\pdfoutput=1

\documentclass[%
 reprint,
 amsmath,amssymb,
 aps, float,
prb,
floatfix
]{revtex4-2}

\usepackage{graphicx, ulem, color, soul, tabularx, textcomp}
\usepackage{dcolumn}
\usepackage{bm, xcolor, siunitx, float}
\begin{document}

\preprint{APS/123-QED}

\title{Higher critical currents yet faster vortex creep in EuBa$_2$Cu$_3$O$_y$ films containing coherent artificial pinning centers}


\author{Jiangteng Liu}
\affiliation{Department of Electrical \& Computer Engineering, University of Washington, Seattle}

\author{Masashi Miura}
\affiliation{Graduate School of Science and Technology, Seikei University, Tokyo, Japan}

\author{Daisaku Yokoe}
\affiliation{Nanostructures Research Laboratory, Japan Fine Ceramics Center, Nagoya, 456-8587, Japan}

\author{Takeharu Kato}
\affiliation{Nanostructures Research Laboratory, Japan Fine Ceramics Center, Nagoya, 456-8587, Japan}

\author{Akira Ibi} 
\affiliation{National Institute of Advanced Industrial Science and Technology, Tsukuba, 305-8564, Japan}

\author{Teruo Izumi}
\affiliation{National Institute of Advanced Industrial Science and Technology, Tsukuba, 305-8564, Japan}
  
\author{Serena Eley}
\affiliation{Department of Electrical \& Computer Engineering, University of Washington, Seattle, WA 98195}

\date{\today}

\begin{abstract}
The electromagnetic properties of type-II superconductors depend on vortices --- magnetic flux lines whose motion introduces dissipation that can be mitigated by pinning from material defects. The material disorder landscape is tuned by the choice of materials growth technique and incorporation of impurities that serve as vortex pinning centers. For example, metal organic deposition (MOD) and pulsed laser deposition (PLD) produce high-quality superconducting films with uncorrelated versus correlated disorder, respectively. Here, we study vortex dynamics in PLD-grown EuBa$_2$Cu$_3$O$_y$ films containing varying concentrations of BaHfO$_3$ inclusions and compare our results with those of MOD-grown (Y,Gd)Ba$_2$Cu$_3$O$_y$ films.  Despite both systems exhibiting behavior consistent with strong pinning theory, which predicts the critical current density $J_c$ based on vortex trapping by randomly distributed spherical inclusions, we find striking differences in the vortex dynamics owing to the correlated versus uncorrelated disorder. Specifically, we find that the EuBa$_2$Cu$_3$O$_y$ films grown without inclusions exhibit surprisingly slow vortex creep, comparable to the slowest creep rates achieved in (Y,Gd)Ba$_2$Cu$_3$O$_y$ films containing high concentrations of BaHfO$_3$. Whereas adding inclusions to (Y,Gd)Ba$_2$Cu$_3$O$_y$ is effective in slowing creep, BaHfO$_3$ increases creep in EuBa$_2$Cu$_3$O$_y$ even while concomitantly improving $J_c$. Lastly, we find evidence of variable range hopping and that $J_c$ is maximized at the BaHfO$_3$ concentration that hosts creep of large vortex bundles or a Bose glass state.

\end{abstract}

\maketitle

\section{\label{sec:introduction} Introduction}

Rare-earth-based superconducting cuprates REBa$_2$Cu$_3$O$_y$ (REBCO) on metallic substrates (coated conductors) have attracted considerable attention for use in high-power, high-magnetic-field applications owing to their ability to carry large, resistance-free currents \cite{Moldyk2023, MacManus-Driscoll2021, Hahn2019, Foltyn2007}.  To carry sufficiently high currents, defects are introduced into these materials to slow the dissipative motion of vortices—quantized magnetic flux lines that penetrate superconductors exposed to magnetic fields, induce noise, create current instabilities, and limit the critical current density $J_c$ \cite{PhysRevB.41.8986, MacManus-Driscoll2021, Kwok2016, Puig2024, Eley2017, Kang2006, Ruiz2026}. Each vortex consists of supercurrents encircling a non-superconducting core of radius $~\xi$ (coherence length). Because vortices can minimize their core energies by pinning to non-superconducting inclusions, these defects are termed \textit{artificial pinning centers} (APCs).  The electromagnetic properties depend on the vortex phases, in which vortices may behave rigidly or elastically, resulting in glassy, plastic, or liquid phases \cite{Blatter1994}. These phases depend on the complex interplay between current-induced forces that propel vortices, the morphology and arrangements of APCs that pin them in energy wells $U$, thermal energy that excites them to hop out of these wells, and vortex elasticity \cite{PhysRev.140.A1197, PhysRevB.41.8986, Eley2021}.

Among various APC candidates, barium-based perovskites Ba\textit{M}O$_3$ (B\textit{M}O, \textit{M} = metal) have demonstrated good lattice matching, chemical stability, and exceptional pinning efficiency \cite{MacManus-Driscoll2004, Haugan2004, Matsumoto2009, Teranishi2008, EleyStrongpinning2021, Gutierrez2007, Xu2014}, with BaHfO$_3$ (BHO) producing the highest $J_c$ performance among tested metal oxides \cite{Miura2017, Tobita2012, EleyStrongpinning2021}. Regarding superconducting hosts, YBCO has long reigned as the prevailing option for the highest currents --- notably, overdoped (Y,Gd)BCO films grown via metal organic deposition/decomposition (MOD) with BHO inclusions have achieved $J_c \approx 32.4\% J_d$, for depairing current density $J_d$ \cite{Miura2022}.  This was a major achievement, considered to be the theoretical ceiling for $J_c$ when vortices are pinned by core pinning \cite{Gurevich2007, Wimbush2015, matsushita2007}. However, this result is limited to 4.2 K and self-field ---  $J_c$ is heavily suppressed at elevated temperatures and magnetic fields, and thermally activated vortex motion (vortex creep) is fast in (Y,Gd)BCO films. For applications under these conditions, other REBCO materials and growth processes may prove superior, which requires understanding vortex phases and dynamics in these candidate materials based on the defect morphologies. Of particular interest, EuBa$_2$Cu$_3$O$_y$ (EuBCO) films grown via pulsed laser deposition (PLD) \cite{Ibi2019, Fujita2019, Ibi2020, Yokoe2020, Wu2022, Lojka2023, Zhao2025, Suzuki2025, ZhaoPo2025} may perform better than (Y,Gd)BCO at high temperatures and magnetic fields \cite{Takahashi2006, Yoshida2014, Yoshida2015}. 

Here, we investigate vortex phases, vortex-defect interactions, and vortex creep in PLD-grown EuBCO-coated conductors doped with BHO at concentrations resulting in volume fractions (vol\%) of 0, 1.9, 2.8, 3.8, and 4.8. Transmission electron microscopy images reveal that the films contain coherent inclusions that transition from nanoparticles at low BHO concentrations to nanorods at higher BHO concentrations. We find that, although the BHO inclusions are effective in boosting $J_c$, with maximum gains achieved at 2.8 vol\%, they speed up vortex creep. This is contrary to what we observe in MOD-grown (Y,Gd)BCO coated conductors containing incoherent BHO inclusions, in which BHO induces concomitant increases in $J_c$ and substantial reductions in creep. We also find that creep in the undoped EuBCO film is surprisingly slow, similar in fact to that in MOD-grown (Y,Gd)BCO containing BHO inclusions. Lastly, we compare our results to strong vortex pinning and Bose glass theories.

\section{Results and Discussion}

In this work, we studied five EuBa$_2$Cu$_3$O$_y$ films that were deposited using PLD and two (Y,Gd)Ba$_2$Cu$_3$O$_y$ films grown using MOD. Each sample contains a different concentration of BHO precursors, and we included reference samples containing no impurities. Table \ref{tab:samples} summarizes sample properties, and the Methods section details the growth process. The oxygen content $y$ is calculated using the empirical relation $0.21y = 1.283 + p$ \cite{Tallon1995}, where $p$ is the carrier concentration from Ref.~\cite{Miura2022}. It is important to note that under similar O$_2$ post-annealing conditions, the carrier concentration $p$ in REBCO films depends strongly on both the choice of rare-earth ion and the presence of artificial pinning centers. Consequently, we used annealing temperatures to achieve near optimal doping in the EuBCO films with versus without BHO impurities \cite{Miura2022}. Therefore, achieving identical doping across all samples is not possible. In this work, we instead maintain doping consistency within each growth method (PLD and MOD).

\begin{table}[h!] 
\caption{\textbf{Sample characteristics.} This table covers the superconductor type, hole concentration $p$, BaHfO$_3$ doping level used in the precursor (mole \%), volume fraction occupied by the BaHfO$_3$ inclusions (vol. \%), approximate inclusion separation $d$, and film critical temperature $T_c$. All EuBCO films were grown by pulsed laser deposition and (Y,Gd)BCO films were grown by metal organic deposition. Inclusion diameters are approximately $4.5 \pm 1 \mathrm{\ nm}$ and 12-14 nm in the EuBCO and (Y,Gd)BCO, respectively. Samples (\#6) and (\#7) also contain a density of 0.03$\times 10^{21}$ m$^{-3}$ 94-nm diameter Y$_2$Cu$_2$O$_5$ (225) precipitates. All films were 550 - 600 nm thick, with the lateral dimensions of $\sim$3 - 4.5 mm.}\label{tab:samples}
\begin{tabularx}{1\linewidth}{clcccc}
\hline \hline
    sample & material & $p$  & mole (vol.) & $d$ & $T_c$\\[-.0em]
    ID  &   & per Cu & \% BHO  & [nm] & [K]\\[-.0em]
    \hline
    1 & EuBa$_2$Cu$_3$O$_{6.87}$ & 0.16 & 0(0) & 0 &  92.8\\
    2 & EuBa$_2$Cu$_3$O$_{6.83}$ & 0.152 & 3.5(1.9) & $30 \pm 10$ &  91.7\\
    3 & EuBa$_2$Cu$_3$O$_{6.83}$ & 0.152 & 5.0(2.8)  & $25 \pm 10$  & 90.5\\
    4 & EuBa$_2$Cu$_3$O$_{6.83}$ & 0.152 & 7.5(3.8) & $20 \pm 5$  & 90.3\\
    5 & EuBa$_2$Cu$_3$O$_{6.83}$ & 0.152 & 10(4.8) & $15 \pm 5$  &  90.2 \\
    6 & (Y,Gd)Ba$_2$Cu$_3$O$_{6.95}$ & 0.177 & 0(0)  & 0 & 92\\
    7 & (Y,Gd)Ba$_2$Cu$_3$O$_{6.97}$ & 0.18 & 15(8.9) & $25 \pm 5$  &  92\\ 
\hline \hline
\end{tabularx}
\end{table}

\begin{figure*}[!t]
\centering
\includegraphics[width=1\linewidth]{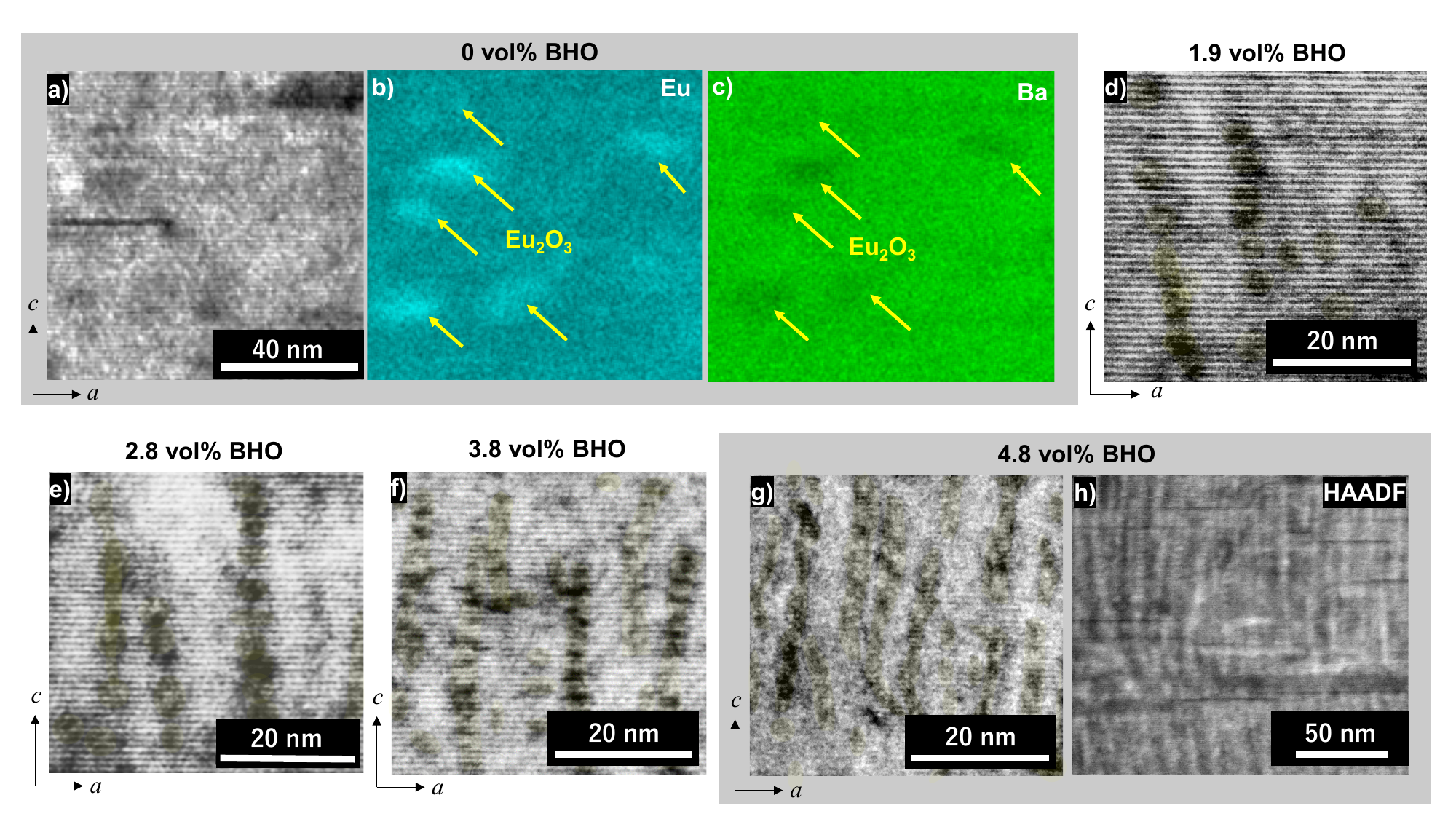}
\caption{\label{fig:fig1} \textbf{Cross-sectional microstructure of EuBCO films.} TEM images of EuBCO films containing (\textbf{a}-\textbf{c}) 0 vol\%, (\textbf{d}) 1.9 vol\%, (\textbf{e}) 2.8 vol\%, (\textbf{f}) 3.8 vol\%, and (\textbf{g}, \textbf{h}) 4.8 vol\% BHO inclusions. Here, (\textbf{b}, \textbf{c}) are EDS for the elements Eu and Ba, respectively, and the yellow arrows identify Eu$_2$O$_3$ precipitates. (\textbf{h}) High-angle annular dark-field (HAADF) STEM image of film containing 4.8 vol\% BHO inclusions.}
\end{figure*}

To determine the average size, shape, and arrangement of the inclusions as well as identify other defects, the films were characterized via scanning electron microscopy (SEM), transmission electron microscopy (TEM), and energy dispersive spectroscopy (EDS). Figure \ref{fig:fig1} shows cross-sectional TEM images of the EuBCO films with different BHO concentrations. The prominent horizontal stripes that appear in all TEM images reveal the CuO$_2$ planes. In the films without BHO, TEM [Fig. \ref{fig:fig1}(a)] and EDS [Fig. \ref{fig:fig1}(b,c)] show $\sim$20-nm-sized Eu$_2$O$_3$ precipitates. 

For films containing BHO, the inclusions are aligned, forming slightly splayed stacks of BHO nanoparticles or nanorods of diameter $4.5 \pm 1 \mathrm{\ nm}$. The nanorod morphology is especially visible in the HAADF image in Fig.\ref{fig:fig1}(h). Note that the nanorod density appears higher than in actuality because the cross-section images capture multiple layers, such that nanorods at different depths appear adjacent. Plan-view HAADF images can be used to estimate the nanorod densities, e.g. The chemcial composition map and HAADF images in Supplementary Fig. S2 show that the BHO nanorods are approximately 20 nm apart in the film containing 4.8 vol\% BHO.

\begin{figure*}[!ht]
\centering
\includegraphics[width=1\linewidth]{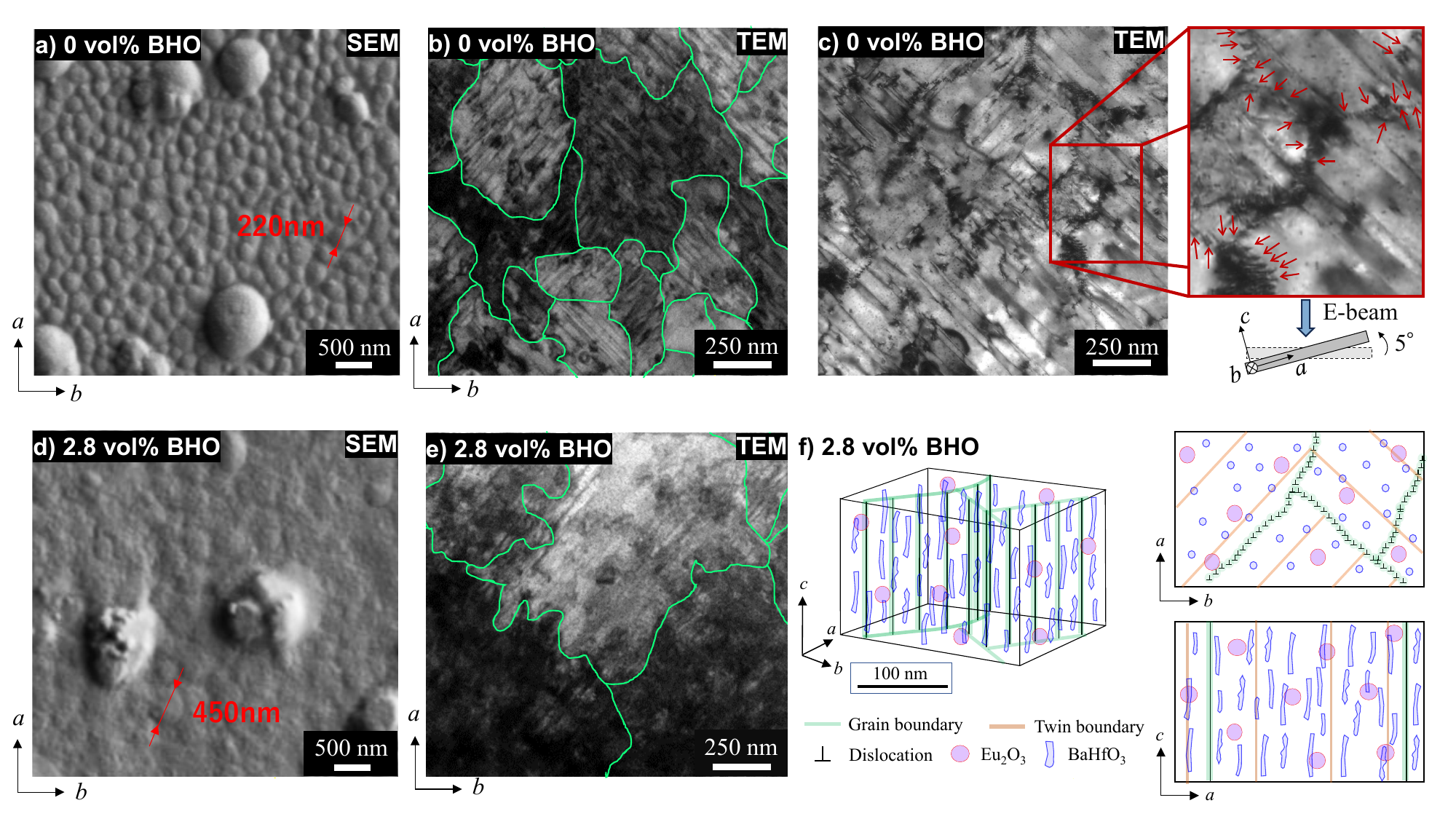}
\caption{\label{fig:fig2} \textbf{Planar microstructure and schematic views of EuBCO films.} (\textbf{a}, \textbf{d}) Planar view SEM images for 0 and 2.8 vol\% BHO sample. (\textbf{b}, \textbf{e}) Planar view TEM images of the two samples; the green lines indicate the grain boundaries. (\textbf{c}) TEM image of the 0 vol\% BHO sample at 5\textdegree tilt to reveal the dislocations at grain boundaries. (\textbf{f}) Illustrations that reveal the microstructure information for the 2.8 vol\% BHO sample. Twin boundaries were not drawn in the 3D view for clarity.}
\end{figure*}

Figure \ref{fig:fig2} presents the plan-view SEM and TEM images of our EuBCO films containing 0 and 2.8 vol\% BHO, highlighting their grain structures. Arrows in the SEM images in Figs. \ref{fig:fig2}(a, d) highlight example grain widths, exemplifying how the average grain size in the film containing BHO is about twice that of the films without BHO.  Consequently, given the smaller grain sizes, the films without BHO contain a higher density of grain boundaries than those with BHO. The corresponding TEM images, Figs. \ref{fig:fig2}(b, e), reveal twin boundaries that appear as striation patterns within individual grains. Here, grain boundaries were identified from the direction of the twin boundaries and the contrast. Fig. \ref{fig:fig2}(c) shows that the grain boundaries in the pristine EuBCO sample contain a high density of dislocations. In PLD-grown REBCO films, these dislocations are predominantly oriented along the $c$-axis \cite{Dam1999, Huijbregtse2000, Miura2006}. To highlight the defect landscape of our EuBCO films, an illustration from three different perspectives: 3D, $ab$-plane slices, and $ca$-plane slices is included in Fig. \ref{fig:fig2}(f).

Regarding the (Y,Gd)BCO films, both contain a sparse density (0.03 $\times$ 10$^{21}$ m$^{-3}$) of 94 nm Y$_2$Cu$_2$O$_5$ precipitates. Sample \#7 also contains 12-14 nm BHO nanoparticles, and the TEM images (see Ref. \cite{Miura2013k}) reveal an increased density of two types of planar defects --- c-axis oriented twin boundaries and stacking faults along the ab-plane --- compared to (Y,Gd)BCO films without APCs. TEM studies \cite{Miura2013k} found that the stacking faults are short ($50\text{-}\SI{100}{\nano\meter}$) and do not segment the twin boundaries, which maintain their integrity throughout the thickness of the film.  The twin boundary spacing is \textasciitilde $\SI{25}{\nano\meter}$ in the film containing BHO and \textasciitilde$\SI{45}{\nano\meter}$ in the reference film (\#7). Comparing the structure of the EuBCO and (Y,Gd)BCO films without impurities, both materials have similar stacking fault and twin boundary densities. However, the EuBCO films have a higher density of grain boundaries and dislocations.

\subsection*{Critical current density}

Analysis of the field-dependent critical current density $J_c(B)$ can reveal the vortex pinning mechanism. Segments of a vortex line or vortex bundles can be pinned by the independent action of large defects (strong pinning) or by the collective action of many small defects (weak pinning) \cite{Larkin1979, Blatter1994, Blatter2004a, Ovchinnikov1991}. Ref. \cite{EleyStrongpinning2021} investigated whether B$M$O nanoparticles act collectively or as strong pinning centers, and revealed consistencies in the $J_c(H)$ data in (Y,Gd)BCO films containing different sizes and densities of nanoparticle inclusions with strong pinning theory.

Notably, simulations based on the time-dependent Ginzburg-Landau (TDGL) theory \cite{Willa2017, Willa2018b, Buchacek2018} predict a power-law relationship between $J_c(B) \propto B^{-\alpha}$, consistent with analytical models \cite{Ovchinnikov1991, Van_der_Beek2002}. Here, the power-law exponent $\alpha$ depends on the size of the spherical inclusions $a$ compared to the in-plane coherence length $\xi_{ab}$, the nanoparticle density $n_p$ compared to the vortex density, and the volume fraction occupied by the inclusions. The simulations find that a landscape consisting of small nanoparticles at low densities only moderately deforms the vortex lattice, whereas an increasing volume fraction results in substantial disordering. For $a \sim 2\xi$, $\alpha$ drops from 0.66 to 0.3 with increasing nanoparticle density. At higher fields, when all inclusions are occupied by a vortex segment, $J_c$ drops faster than $B^{-1}$ due to the weakening of pin-breaking forces at higher fields. Lastly, it is important to note that the TDGL simulations are based on monodispersed spherical inclusions that may accurately represent the pinning environment in MOD-grown samples with uncorrelated disorder. Here, we evaluate whether the results of these simulations are also accurate in the case of correlated disorder found in our PLD-grown samples.

\begin{figure*}[!ht]
\centering
\includegraphics[width=0.75\linewidth]{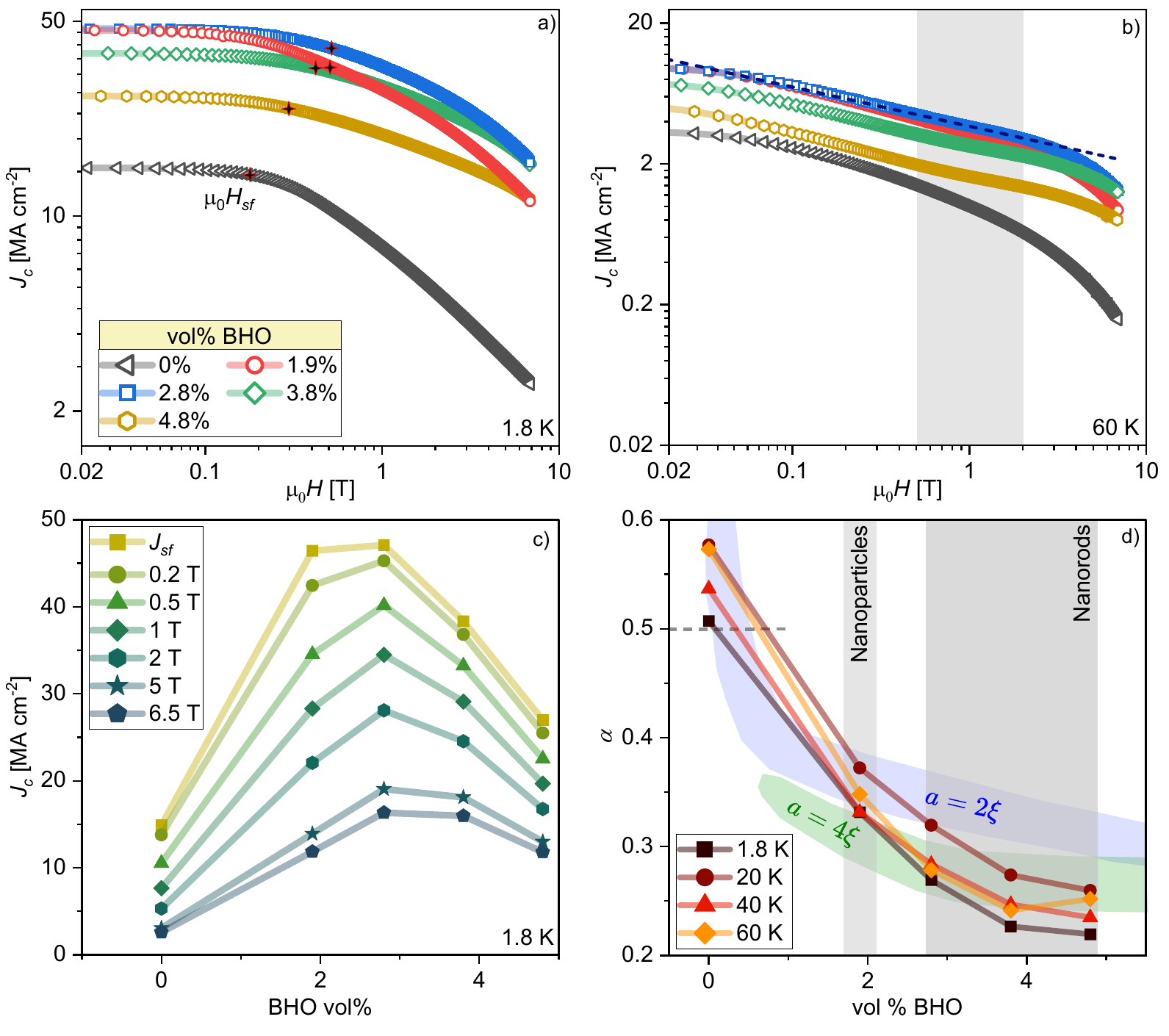}
\caption{\label{fig:fig3} \textbf{BHO doping dependence of $J_c$ and pinning behavior in EuBCO films.} Field-dependent $J_c$ for EuBCO films with different BHO doping levels at a temperature of (\textbf{a}) 1.8 K and (\textbf{b}) 60 K. The compass stars in (\textbf{a}) mark the calculated self-field $H_{sf}$. (\textbf{c}) Critical current density $J_c$ versus vol\% BHO at different applied magnetic fields with temperature at 1.8 K. (\textbf{d}) Comparison of extracted $\alpha$ vs vol\% BHO with predictions from TDGL simulations based on strong pinning theory \cite{Willa2017} at different temperatures. The colored shaded region represents simulation predictions for nanoparticles of diameter $a$ equivalent to twice (blue) and four times (green) the coherence length $\xi_{ab}(T)$. The power-law exponent \(\alpha\) is extracted from the slope of a linear fit to $\log J_c - \log \mu_0H$, restricted to a field region in which \(J_c\propto B^{- \alpha}\). The dashed line in (\textbf{b}) shows an example fit applied within the shaded region (0.55 - 2 T). In (\textbf{d}), the dashed line at $\alpha=0.5$ is the prediction when planar dislocations pin vortices. }
\end{figure*}

To determine $J_c(B)$ in our EuBCO films, we performed magnetization studies in a Quantum Design MPMS3 magnetometer. Specifically, we collected magnetic hysteresis loops $m(H)$ at fixed temperatures, then applied the Bean critical state model for rectangular samples to convert $m(T,H)$ into $J_c(T,H)$ \cite{Gyorgy1989, Talantsev2024BeanModel}. Accordingly, we calculated $J_c(T,H) = 20\Delta m(T,H)/[w^2l\delta(1-w/3l)]$, where $\Delta m$ represents the difference between the magnetic moments $m$ in upper and lower branches of the magnetic hysteresis loops at a given field, whereas $w$, $l$, and $\delta$ denote the widths, lengths, and thicknesses of the samples. The critical current density $J_c$, magnetic moment $m$, sample dimensions, and coefficient 20 have units of A cm$^{-2}$, emu, cm, and A cm$^{-2}$ emu$^{-1}$, respectively \cite{Talantsev2024BeanModel}. The magnetization loops are provided in Supplementary Fig. S3, and further measurement details are explained in the Methods section.

Figures~\ref{fig:fig3}(a) and (b) display the calculated $J_c$ as a function of the applied magnetic field, plotted on a logarithmic scale at 1.8 K and 60 K, respectively, while the corresponding data at 20 K and 40 K are provided in Supplementary Fig. S4. From these plots, distinct regions of differing vortex behavior are apparent. At low magnetic fields, $J_c$ remains nearly constant with increasing field, characterizing the self-field regime $\mu_0H_{sf}=\mu_0\gamma J_{sf}\delta/\pi$ in film,\cite{Polat2011} where $\gamma$ is the critical field anisotropy and $J_{sf}$ is the critical current density at zero field. Using $\gamma = $ 5 \cite{Tsuchiya2017, Bartolomé2019}, we calculate $\mu_0H_{sf} = $ $0.179-0.518$ T for all samples and identify self-field with crosses in Fig.~\ref{fig:fig3}(a).  Above self-field, $J_c$ follows the power-law dependence $J_c \propto B^{-\alpha}$ expected in the case of strong pinning theory, and observed in (Y,Gd)BCO films and other superconductors containing B$M$O inclusions \cite{Maiorov2009,EleyStrongpinning2021,Polat2011,Ijaduola2012}. Finally, as the applied magnetic field approaches the irreversibility field, most visible in the high-field 60 K data, $J_c$ drops more rapidly. For information regarding the temperature-dependent $J_c$ is included in Supplementary Fig. S5(c).


Next, we examine the variation of $J_c$ with BHO concentration at 1.8 K under different applied magnetic fields, as shown in Fig. \ref{fig:fig3}(c). In all fields, BHO inclusions enhance $J_c$, with a peak at 2.8 vol\%. At self-field and 5 T, the critical current density increases by factors of 3.2 and 6.2, respectively, relative to the undoped sample. These trends align with predictions from the TDGL model, which suggests an optimal inclusion density exists when $a=4
\xi$. Furthermore, Fig. \ref{fig:fig3}(c) indicates that this optimal density shifts to higher BHO concentrations with increasing magnetic field. Specifically, for fields above 1 T, the sample with 3.8 vol\% BHO outperforms that with 1.9 vol\%, and may surpass the 2.8 vol\% sample at fields beyond the measured range, as shown in Fig.~\ref{fig:fig3}(a). This field-dependent shift in the optimal inclusion density is also consistent with TDGL model predictions \cite{Willa2017}. 
 
To understand the pinning mechanism, we extract the exponent $\alpha$ from $J_c(B)$, and plot it in Fig.~\ref{fig:fig3}(d) versus BHO concentration. First, considering the film containing no inclusions, we find that $\alpha \approx 0.5$ at 1.8 K, indicative of vortex pinning by dislocations \cite{Diaz1998, Gurevich1994}. This is consistent with our microscopy results suggesting relatively high dislocation densities in these films. At higher temperatures, the vortex core size increases, weakening the effectiveness of dislocations (owing to their minuscule size) as pinning sites, evidenced by an increase in $\alpha$. By comparison, the (Y,Gd)BCO films without BHO, which hosts fewer dislocations, exhibit $\alpha = 0.65$ at low temperatures \cite{EleyStrongpinning2021}. Adding BHO dramatically decreases $\alpha$, unsurprisingly indicating the insignificance of pinning by dislocations in the presence of strong pinning centers. With increasing BHO content, $\alpha$ decreases from 0.58 to 0.22. To compare these results with the strong pinning theory formulism, the blue and green shaded regions represent the $\alpha$ values predicted by TDGL simulations for the inclusion diameter of $a = 2\xi$ and $a = 4\xi$, respectively. Assuming that the coherence length is similar to that in YBCO --- $\xi_{ab}(0) \approx 2 \mathrm{\ nm}$ \cite{Abou_El_Hassan2021}--- we estimate $\xi_{ab}(T) =\xi_{ab}(0)/(1-T/T_c)^{1/2} \approx 2-3.4 \mathrm{\ nm}$  for $1.8 - 60$ K, and consider the particle/rod diameter of $4.5$ nm. 
Our measured $\alpha$ matches the $a = 2\xi$ case at 1.9 vol\% and transitions to the $a = 4\xi$ prediction at higher dopings of 2.8, 3.8, and 4.8 vol\%. This trend can be explained by a morphological change from nanoparticles to nanorods, in which the nanorods may be considered to have a larger equivalent radius.

\subsection*{Magnetic relaxation as a result of vortex creep}

Vortex creep studies are a useful tool for determining the vortex pinning energy barriers, vortex configurations, ordering, and dynamics. To reduce their core energies by a pinning energy $U_0$, vortices localize in potential energy wells defined by material defects. This results in energy barriers that must be surpassed for vortices to move through the material. Currents slant this energy landscape, lowering the energy barrier to a current-dependent function $U(J)$. Vortices can hop out of these energy wells, due to thermal energy (vortex creep), and the resulting dissipation from vortex motion lowers the induced current from the critical current $J_{c0}$ to $J$.

Different models propose different relationships for $U(J)$. For example, the Anderson-Kim model is an early formulism that approximated $U(J)=U_0(1-J/J_{c0})$ by treating vortices as rigid rods and disregarding vortex-vortex interactions, which are most significant at high magnetic fields \cite{Yeshurun1996, Blatter1994}. This model can be fairly accurate at low temperatures ($T \ll T_c$) and low magnetic fields, as well as in the early stages of the relaxation process ($J \lesssim J_{c0}$), where $J_{c0}$ is the initial induced current, assumed to be close to the critical current $J_c$. In the later stages of relaxation in which $J \ll J_{c0}$, vortex elasticity becomes important, such that collective creep theories predict an inverse power law form for the energy barrier $U(J)=U_0[(J_{c0} / J)^\mu]$ \cite{Feigelman1989, Blatter1994}. Here, the glassy exponent $\mu$ relates to the size of the vortex bundle that jumps during the creep process.  To capture behavior for a broad range of $J$, an interpolation formula between the two regimes is commonly used:
\begin{align}\label{eq:Uinterpolation}
U(J)=U_0 [(J_{c0}/J)^\mu -1] / \mu. 
\end{align}
\noindent Considering Eq.~\eqref{eq:Uinterpolation} and the Arrhenius relation for the time required for thermal activation over the barrier
\begin{align}\label{eq:creeptime}
t=t_0 e^{U(J)/k_BT}, 
\end{align}
one finds the expected decay in the persistent current over time $J(t)$ and, subsequently, the vortex creep rate $S(T)$:\cite{Blatter1994} 
\begin{align}\label{eq:Jtdecay}
J(t) \propto M(t)=M_{0} \Big[1+\frac{\mu k_B T}{U_0} \ln (t/t_0)\Big]^{-1/\mu}
\end{align}
\begin{align}\label{eq:ST} S \equiv \left| \frac{d \ln M}{d \ln t} \right| = \frac{k_B T}{U_0+\mu k_B T \ln (t/t_0)}. 
\end{align}
Here, $t_0$ is often referred to as the microscopic time for pinned vortex and related to the attempt frequency, which is typically $\sim10^{-8}-10^{-6}$ s \cite{Blatter1994, blatter_vortex_2003, Kwok2016}.
 
 As evident in Eq.\ \eqref{eq:ST}, creep data provides access to both $U_0$ and $\mu$.  Because $m(t) \propto J(t)$, we can measure creep through magnetic relaxation studies, repeatedly measuring the magnetic moment $m(t)$ for one hour to capture the decay in the magnetization $M(t)$ over time. We then plot this on a logarithmic scale ($\log m - \log t$), and extract $S$ from the slope, according to Eq. (\ref{eq:ST}). Further information regarding our magnetic relaxation measurement and $S$ extraction protocols is detailed in the Methods section.

\begin{figure*}[!ht]
\centering
\includegraphics[width=1\linewidth]{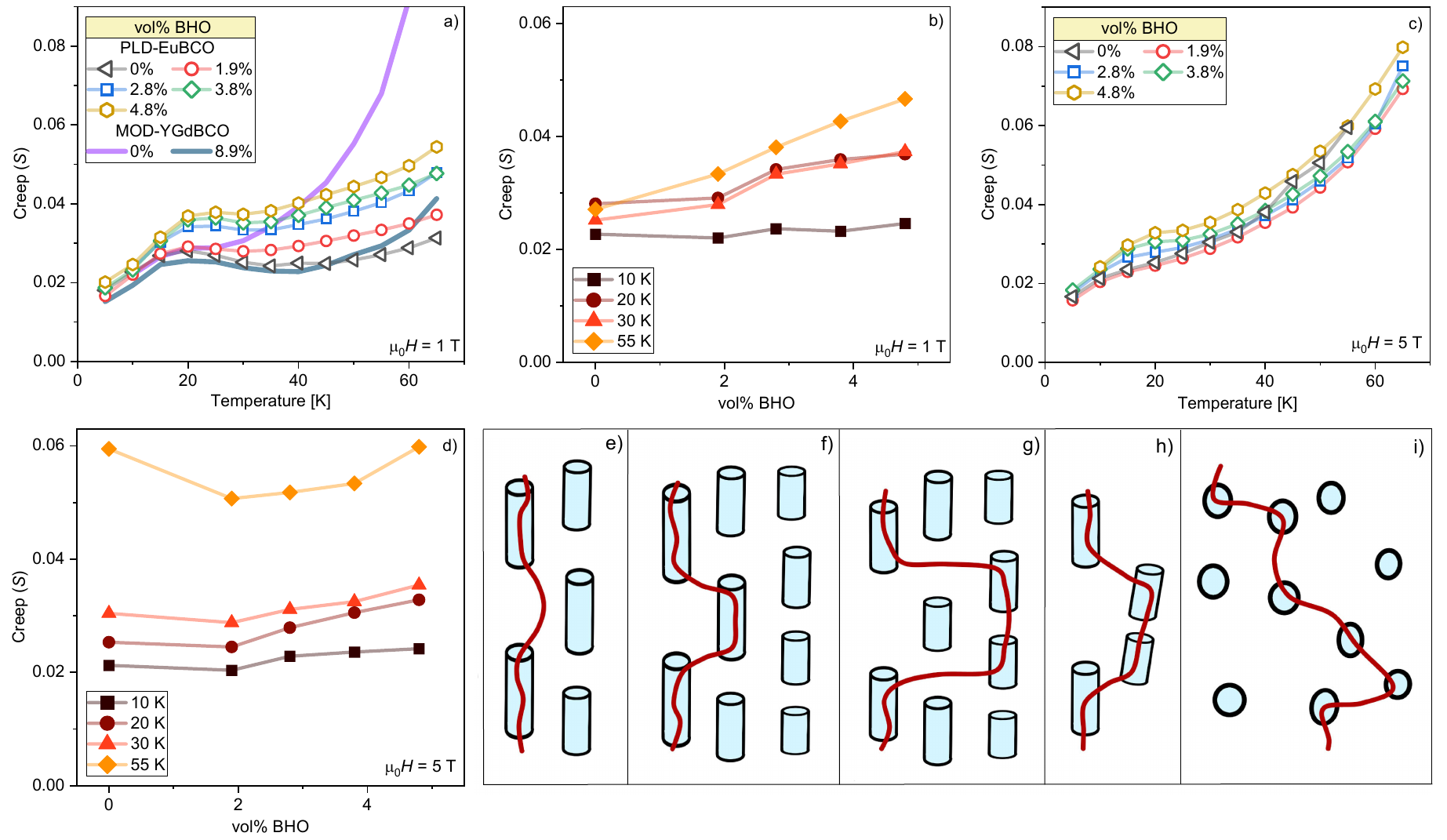}
\caption{\label{fig:fig4} \textbf{Vortex creep behavior of EuBCO films at $\mu_0H = $ 1 and 5 T with pinning schematics.} (\textbf{a, c}) Vortex creep parameter $S$ vs temperature for PLD-grown EuBCO films with different BHO concentrations at $\mu_0H = 1$ and 5 T, respectively. For comparison, data for MOD-grown (Y,Gd)BCO film are also included as solid lines. (\textbf{b, d}) $S$ versus BHO doping at four different temperatures under an applied field of 1 and 5 T , respectively. (\textbf{e}-\textbf{i}) Possible vortex (red line) configurations in a landscape of nanorods (blue cylinders) and nanoparticles (blue circles), which includes (\textbf{e}) half-loop excitations, (\textbf{f}) double-kink excitations, (\textbf{g}) variable-range hopping, (\textbf{h}) splay glass, and (\textbf{i}) strong pinning by nanoparticles.}
\end{figure*}

Angle-dependent $J_c(\theta)$ measurements in REBCO films have shown that different types of defects dominate vortex pinning depending on temperature. At low temperatures ($T < 20$~K), point-like pinning centers such as vacancies, intergrowths, and nanoparticles provide the main contribution to vortex pinning. In contrast, at higher temperatures, columnar and planar objects such as twin boundaries, grain boundaries, and nanorods become more effective pinning centers. This temperature-dependent crossover between isotropic and anisotropic pinning mechanisms has been widely observed in REBCO systems \cite{Tsuchiya2017, Palau2018, Puig2024}. Figure~\ref{fig:fig4}(a) compares the temperature dependence of $S$ at 1 T in our samples. First, notice that at low temperatures $T \lesssim 15$ K, creep increases roughly linearly, consistent with $\mu k_B T \ln(t/t_0) \ll U_0$ leading to $S \sim k_BT/U_0$, as predicted by the Anderson-Kim model \cite{Blatter1994, Eley2017}. Moreover, there is little variation in $S$ with BHO concentration in this regime, indicating that at these temperatures, vortex elasticity remains insignificant and the pinning mostly comes from isotropic point defects instead of the BHO nanorod, consistent with other studies \cite{EleyStrongpinning2021, Eley_2017a, Tsuchiya2017, Palau2018, Puig2024}.  This is exemplified in Fig. \ref{fig:fig4}(b), showing that $S$ at 10 K is relatively insensitive to the volume percentage of BHO.

At higher temperatures $T \gtrsim 20 \mathrm{\ K}$, the anisotropic pinning from BHO inclusions strongly affect vortex creep. In the (Y,Gd)BCO film, BHO consistently reduces $S$ at all temperatures, though this decrease is far more pronounced at higher temperatures.  This is evident by comparing the solid curves in Fig. \ref{fig:fig4}(a) and in other studies of (Y,Gd)BCO films containing BHO \cite{EleyStrongpinning2021, PhysRevB.83.184519, Miura2017}. In contrast, however, in the EuBCO films, $S$ increases with BHO concentration [see Figs.  \ref{fig:fig4}(a) and (b)] A surprising result in Fig. \ref{fig:fig4}(a) is that the EuBCO sample without BHO shows slow creep behavior that is remarkably similar to that in the 8.9 vol\% BHO-doped YGdBCO. Hence, the intrinsic disorder inherent to the PLD-growth process is far more effective in slowing creep than that produced by MOD. Similar trends where $S$ is lower for the undoped sample are also observed in PLD-grown YBCO films containing B\textit{M}O nanorods. More information is included in the Supplementary  Fig. S8, where we compare our EuBCO results with YBCO data from Ref. \cite{Mele2008, Maiorov2009}.

Creep in the EuBCO films collected at 1 T becomes relatively insensitive to temperature above 20 K. This plateau is particularly pronounced in the samples containing low BHO concentrations and exists over a narrower temperature range in the more concentrated samples. Plateaus may appear in $S(T)$ when $\mu k_BT \ln(t/t_0) \gg U_0$ such that $S \approx [\mu \ln(t/t_0)]^{-1}$ (see Eq. \ref{eq:ST}), and are often associated with glassy behavior \cite{Eley2018, Eley2020sr, Yeshurun1996, PhysRevB.42.6784,Kwok2016}. Originally described by Nelson and Vinokur, glassy phases are pinning-induced metastable states in which vortices become localized, resulting in strongly enhanced vortex pinning \cite{Nelson1992, Nelson1993, vanderBeekHandbook}. In fact, in the absence of a current $J \rightarrow 0$, the creep barrier diverges $U(J) \rightarrow \infty$ resulting in zero resistivity $\rho \propto e^{-U(J)/k_BT}$.

Next, we consider creep at higher vortex densities \cite{Kwok2016}. Figure~\ref{fig:fig4}(c) displays $S(T)$ in a field of 5 T and Fig.~\ref{fig:fig4}(d) highlights how $S$ depends on BHO concentration. From both figures, we see that $S$ is less sensitive to BHO concentration than at 1 T. Though $S$ still generally increases with BHO concentration, the sample containing 1.9 vol\% BHO now demonstrates the slowest creep rather than the undoped sample, as was the case at 1 T.  This suggests that intrinsic lattice defects in pure EuBCO provide effective pinning at lower fields, but their density is insufficient to pin all of the vortices at higher fields. Additionally, a clear increase in $S$ occurs at doping levels above 1.9 vol\%. Additional creep data collected at other magnetic fields is included in Supplementary Fig. S6.

We can assess the type of glassy dynamics in our sample by extracting $\mu$ and comparing it to predictions based on collective creep and Bose glass theories. According to collective creep theory, the exponent $\mu$ depends on whether a single vortex or a vortex bundle of lateral dimensions smaller than (small bundle) or larger than (large bundle) the penetration depth $\lambda_{ab}$ hops due to thermal activation. Given the relatively low anisotropy of EuBCO, we focus exclusively on the 3D collective creep regime, in which case $\mu=1/7$ is expected for hopping of single vortices, 3/2 for small bundles of flux, and 7/9 for large bundles of flux \cite{Blatter1994}. Note that these predictions do not consider a mixed vortex pinning landscape composed of different types of defects, which may modify the calculated exponents.

Bose glass theory \cite{Nelson1992, Nelson1993} describes a variety of vortex excitations associated with a Bose-glass state in a landscape of correlated disorder, illustrated in Fig. \ref{fig:fig4}(e-i). Starting with a flux line pinned to a linear defect (or alignment of defects) with inhomogeneous pinning energy, forces from a current may excite a curved segment called a half-loop in which the barrier $U \sim 1/J$ ($\mu = 1$) \cite{Kwok2016, Nelson1992}. These excitations typically appear at relatively low temperatures and fields, and in the early stages of the magnetic relaxation process ($J \lesssim J_{c0}$). At low fields and increasing temperatures, the loops expand in size and may pin to a neighboring defect, forming a double-kink excitation that characterizes a non-glassy phase in which the kinks slide rapidly, increasing creep. Otherwise, it may pin to a remote defect having lower energy than an adjacent one, behavior that is reminiscent of Mott variable-range hopping in disordered semiconductors, thus deemed the variable-range hopping (VRH) regime, characterized by a barrier $U \sim (1/J)^{1/3}$ in the case of parallel defects. 

In the VRH regime, the exponent $\mu = 1/3$ is applicable to the case of parallel defects and short-range repulsive interactions between flux lines. Double-kink excitations pinned by splayed defects result in a $\mu$-value whose value is sensitive to the angle between the tracks \cite{Civale1997, Hwa1993}. (For example, it was found that a Gaussian angular distribution of tracks produces $\mu = 3/5$) \cite{Civale1997, Hwa1993}.  Moreover, $\mu$ varies between $1/3$ and $1$ in the case of long-range repulsive vortex-vortex interactions \cite{PhysRevLett.78.4845, PhysRevB.58.6565}, with the specific value depending on the ratio of the penetration depth to the mean pin separation $d$ and the filling factor $f = B/B_\Phi$ (pin occupancy, where $B_\Phi=2\Phi_0/(d^2\sqrt{3})$ is the matching field for a triangular vortex lattice).

\begin{figure*}[!ht]
\centering
\includegraphics[width=1\linewidth]{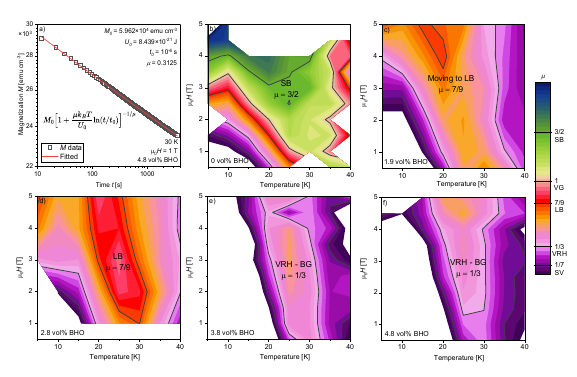}
\caption{\label{fig:fig5} \textbf{BHO doping dependence of vortex phase diagrams in EuBCO films.} (\textbf{a}) Example of glassy exponent $\mu$ extraction by fitting the magnetic relaxation data $M(t)$ using Eq. \eqref{eq:Jtdecay}. Vortex phase diagram of EuBCO films containing (\textbf{b}) 0 vol\%, (\textbf{c}) 1.9 vol\%, (\textbf{d}) 2.8 vol\%, (\textbf{e}) 3.8 vol\%, and (\textbf{f}) 4.8 vol\% BHO inclusions. The contour lines and the marks on the color bar indicate the expected $\mu$ values associated with hopping in the single vortex (SV), small-bundle (SB), and large-bundle (LB) vortex creep regimes, as well as vortex glass (VG) and variable-range hopping (VRH) in Bose glass states.}
\end{figure*}

In the EuBCO films, we find evidence of transitions between creep of different vortex bundle sizes as well as of the Bose glass state from consistency between the experimentally extracted glassy exponents $\mu$ and expected theoretical values. To extract the $\mu$, we fitted the magnetic relaxation data $M(t)$ using Eq.~(\ref{eq:Jtdecay}). Here, $M_0$, $U_0$, and $\mu$ were treated as free parameters, while $t_0$ was fixed at $10^{-6}$~s. It is important to note that while the fitted $M_0$ and $U_0$ are highly dependent on the choice of $t_0$, the extracted $\mu$ remains robust. A more accurate estimate of $t_0$ can be obtained by considering the electric field generated during the magnetic field sweep \cite{Gurevich1993a, Gurevich1994b}. Further details on how $t_0$ affects $M_0$, $U_0$, and $\mu$ are provided in Supplementary Fig.~S9. Figure~\ref{fig:fig5}(a) shows an example fit to the data from the EuBCO film containing 4.8~vol\% BHO measured under 1 T and 30 K, plotted on a logarithmic scale. The fitted $M_0$ and $U_0$ values for other temperatures, fields, and BHO concentrations are provided in Supplementary Fig.~S7. Figure~\ref{fig:fig5}(b–f) plots the extracted $\mu$ for the EuBCO films containing different BHO concentrations as contour maps, where each distinct color represents the predicted $\mu$ values. At lower temperature $T<20$ K, where pinning is highly affected by isotropic defects \cite{Tsuchiya2017, Palau2018, Puig2024}, we observe that $\mu$ approaches approximately 1/7 for all samples, as indicated by the dark purple region. This behavior is consistent with creep governed by single-vortex hopping. The effect is particularly pronounced in samples containing $0-2.8$ vol\% BHO under magnetic fields $\mu_0H < 3$ T.

Next, we focus on temperature between $20\lesssim T\lesssim 35$ K, where anisotropic pinning starts to dominate \cite{Tsuchiya2017, Palau2018, Puig2024}. For the EuBCO sample without BHO (\#1), we see that $\mu \approx 3/2$, indicated by the green color in Fig.~\ref{fig:fig5}(b). This $\mu$ value is commonly observed in YBCO \cite{Cornejo2025, PhysRevB.85.174504, PhysRevB.50.7188} and attributed to creep of small bundles of vortices. For the films containing BHO inclusions, we see that $\mu$ is only weakly field dependent. 

Comparing samples \#2 and \#3, we see that as BHO concentration increases from 1.9 to 2.8 vol\% BHO, $\mu$ approaches 7/9, indicated by the orange-to-red color scale transition in Fig.~\ref{fig:fig5}(c-d). This is consistent with creep of large bundles of vortices or the Bose glass state. To evaluate the possibility of a Bose glass state in these films that consist of half-loops at 1 T and 20 K --- onset of the plateau in $S(T)$ --- we compare the scale of $U^*$ to the theoretical approximation for the pinning energy associated with half-loops $U_{hl} \sim \varepsilon_r \ell_{hl}$ \cite{Blatter1994}. Here, $U^* \equiv U_0+\mu k_BT\ln(t/t_0)$ is the auxiliary energy scale (see Eq. \ref{eq:ST}) \cite{Zhou2016a, Sun_2015, Sun2015d,Haberkorn2011b,Miu2013,Sundar2017} and $\ell_{hl} \sim \xi_{ab}(\varepsilon_r \varepsilon_\ell/\varepsilon_0^2)^{1/2}(J_d/J_c)$ is the critical radius attained by the half-loops that precipitates depinning, $J_d = \Phi_0/(3^{3/2}\pi \mu_0 \lambda_{ab}^2\xi_{ab})$ is the depairing current density \cite{Blatter1994}, $\lambda_{ab}(T)$ is the penetration depth, {$\xi_{ab}(T)=\xi_{ab}(0)/[1-(T/T_c)]^{1/2}$ is the coherence length, $\varepsilon_r$ is the pinning energy per unit length, $\varepsilon_0=\Phi_0^2/(4\pi\mu_0\lambda_{ab}^2)$ is the line energy, and $\varepsilon_l=(\varepsilon_0/\gamma^2)\ln(\lambda_{ab}/\xi_{ab})$ is the line tension (in the nondispersive limit and disregarding anisotropy factors) \cite{Nelson1993}.

Under ideal pinning conditions, $\varepsilon_r \approx \varepsilon_0(R/2\xi_{ab})^2$ when the defect radius $R < \sqrt{2}\xi_{ab}$ (applicable here since $R=a/2 \approx \xi$). Assuming $\xi_{ab}(0) \approx 2 \mathrm{\ nm}$ and $\lambda_{ab}(0) \approx 150 \mathrm{\ nm}$ (similar to that in YBCO \cite{PhysRevApplied.11.014035, PhysRevLett.72.744, PhysRevLett.74.1008, Abou_El_Hassan2021})--- we use $\xi_{ab}({20} \mathrm{\ K}) = 2.26 \mathrm{\ nm}$ and $\lambda_{ab}(20 \mathrm{\ K}) = 158 \mathrm{\ nm}$ \cite{Hardy1993,Prozorov2006} to find the following parameters at $T= 20\mathrm{\ K}$: $J_d \approx 179 \mathrm{\ MA \ cm^{-2}}$, $\varepsilon_0 \approx 1.08\times10^{-11} \mathrm{\ J \ m^{-1}}$, $\varepsilon_r \approx 2.71\times10^{-12} \mathrm{\ J \ m^{-1}}$, and $\varepsilon_l \approx 1.84\times10^{-12}\mathrm{\ J \ m^{-1}}$. Based on these calculations, we estimate that at 20 K $U_{hl}/k_B$ ranges from 760 K (1.9\% BHO) to 527 K (2.8\% BHO). The calculation considers $J_c$-values of $21.5\mathrm{\ MA \ cm^{-2}}$ and $31.0\mathrm{\ MA \ cm^{-2}}$ for 1.9 vol\% BHO and 2.8\% BHO, respectively, extracted from the fitted $M_0$ data. (As a note of caution, the calculated values are highly sensitive to the choice of penetration depth, though this does not change the magnitude of the calculations.) We now compare our theoretical estimates of pinning energy in half loop $U_{hl}/k_B$ to our experimentally extracted $U^*/k_B$ values in the the films containing $1.9$ and $2.8$ vol\% BHO at 1 T and 20 K. Our measured effective pinning barrier is $U^*/k_B = 698-598$ K, using the fitted $U_0 \ (7.43 - 5.85 \times 10^{-21}$ J), $\mu \ (0.364 - 0.395)$, and $t=3600$ s (1h creep). This is indeed comparable to the theoretical estimates predicted for the pinning energies associated with half-loops at 20 K. For films with even higher BHO concentrations, sample \#4 and \#5 (3.8-4.8 vol\%), we see that $\mu \approx 1/3$, indicated by the pale purple color in Fig.~\ref{fig:fig5}(e-f). So the behavior of both samples is consistent with the predictions of variable-range-hopping with long-range vortex-vortex interactions \cite{ Nelson1992, Nelson1993, Civale1997, Hwa1993, PhysRevLett.78.4845, PhysRevB.58.6565, Thompson1997}.

\section*{Conclusions}

In summary, this study compares vortex dynamics in a landscape of uncorrelated versus correlated disorder in otherwise similar systems. By performing magnetization studies, we measure the critical current densities and rates of thermally activated vortex motion in EuBCO films with correlated disorder and compare results to measurements of (Y,Gd)BCO films with uncorrelated disorder conducted in this and previous studies. In the (Y,Gd)BCO films, adding several-nm-sized BHO inclusions slows creep and boosts $J_c$. Slowing creep is congruent with increasing $J_c$; creep leads to a rapid drop in magnetically induced supercurrents and rounding in the current-voltage characteristics under applied current, which is directly responsible for decreased $J_c$.  However, in the EuBCO films, we found two interesting effects: (1) though adding BHO inclusions indeed increases $J_c$, it also quickens creep and (2) the EuBCO film without inclusions exhibits remarkably slow creep.

The increase in creep is not entirely surprising, given that this is a known effect of adding columnar defects, due to the formation of double-kink excitations that expand rapidly, leading to quick vortex hopping between pinning sites. These kinks can be arrested by introducing random defects, which should increase $J_c$. This study therefore motivates exploring whether adding random disorder in PLD-grown EuBCO films containing around 2-4 vol\% BHO may lead to reduced creep and further increases in $J_c$.

\section*{Methods}\label{sec:Methods}

\subsection*{Sample Preparation}

EuBCO films were deposited using pulsed layer deposition onto heated substrates consisting of textured CeO$_2$ (300 - 700 nm), LaMnO$_3$ (7 nm), MgO (5 nm) deposited via ion-beam-assisted deposition (IBAD), Y$_2$O$_3$ (14 nm), Gd$_2$Zr$_2$O$_7$ (56 nm), and Hastelloy C276 (100 nm). The films containing BHO nanoparticles were deposited at
1055 \textdegree C, whereas a temperature of 1150 \textdegree C was used for samples without inclusions. Post annealing, EuBCO films with BHO inclusions were annealed in O$_2$ at 250 \textdegree C for 3 hours. While the 0 vol\% EuBCO was annealed in O$_2$ at 350 \textdegree C for 3 hours. More details have been published elsewhere \cite{Ibi2020}.

The (Y,Gd)123 films were grown epitaxially on buffered tape via metal organic deposition using solutions containing Y-, Gd-, and Ba-trifluoroacetates and Cu-naphthenate (with a cation ratio of 0.77:0.23:1.5:3). To intercalate BHO inclusions, we incorporated Hf-naphthenate (at the indicated volume percentage) into the precursor solutions. The diameter of the BHO nanoparticle is controlled by the thickness of the precursor layer during the coating/calcination steps \cite{Miura2017}. The coating layer thickness used here is 30 nm. The buffered tape contained a stack of CeO$_2$/Y$_2$O$_3$/LaMnO$_3$/IBAD-MgO/Gd$_2$Zr$_2$O$_7$/Hastelloy C276. The overdoped state in (Y,Gd)123 films was achieved through post annealing in O$_2$ at 300 \textdegree C for 3 hours. Ref. \cite{Miura2022} provides further information regarding this procedure.

\subsection*{Magnetometry Measurements}

Magnetization measurements were collected using a Quantum Design MPMS3 superconducting quantum interference device (SQUID) magnetometer. All superconducting films were mounted on delrin disks centered inside a straw. For all measurements, the magnetic field was applied perpendicular to the film plane (parallel to the film's c-axis $H || c$), and each measurement of the moment used a 30 mm scan collected over a period of 4 s. For the film containing 2.8 vol\%, a scan length of 10 mm collected over 1 s was used due to large signals. Lastly, a vibrating sample magnetometry (VSM) mode is used for the $J_c$ result shown in Fig.~\ref{fig:fig3} to reduce the effect of creep. DC scan mode and VSM mode comparisons are shown in Supplementary Fig. S3 and Fig. S4.

To determine the critical temperature $T_c$,  we measured the moment $m(T)$ in an applied field of $\mu_0 H = 0.2-0.4 \mathrm{\ mT}$ while sweeping the temperature at a rate of approximately 0.5 - 4 K min$^{-1}$. Details on $T_c$ extraction are included in Supplementary Fig. S1. Magnetic hysteresis loops $m(H)$ were collected by measuring the magnetic moment stabilized at each field after sweeping at a rate of 100 Oe s$^{-1}$, for the film with 2.8 vol\% BHO, a continuous sweep rate of 100 Oe s$^{-1}$ was used.

Magnetic relaxation (creep) was captured using conventional protocols \cite{Yeshurun1996} of recording the moment $m(t,H_1,T_1)$ every 4-10 s for approximately one hour, after establishing the critical state. To establish the critical state, the field was swept $\Delta H = 2 \ \mathrm{\ T} > 4H^*$, where $H^*$ is the minimum field at which magnetic flux fully penetrates the sample, then fixed at the field $H_1$ of interest. For each relaxation measurement, we also fix at an intended temperature $T_1$. Establishment of the critical state was corroborated by comparing $m(t)$ to the magnetic hysteresis loops $m(H)$, verifying that the initial measurement does indeed lie on the loop. After subtracting the background produced by the sample mount and adjusting the time to account for the difference between the initial application of the field and the first measurement (maximizing the correlation coefficient), $S = -d\ln m/d\ln t$ is extracted from the slope of a linear fit to $\ln m$ versus $\ln t$. More information on how relaxation data is processed is provided in Supplementary Fig. S5. Lastly, the magnetization $M$ is calculated using the moment $m$ and the sample volumes, where the exact sample dimensions are in Supplementary Table S1. 

\section*{Data availability}

The data supporting the findings of this study are available on Mendeley Data (doi.org/10.17632/z7t5jtyvmw.2) as a zip file. This includes Origin files (.opju) that contain raw and processed data spreadsheets for all the samples and figures used in this paper, which can be opened using Origin Viewer, a free application that permits viewing and copying of data contained in Origin project files. README files are also included to guide the reader about the contents of each folder and provide captions and explanations for the corresponding data files.

\section*{Code availability}

The custom Python codes used to process the raw data are available on Mendeley Data (doi.org/10.17632/z7t5jtyvmw.2) as a zip file. README files are included to guide the reader about the contents of each folder and provide captions and explanations for the corresponding data files. JupyterLab version 4.0.10 and Python version 3.11.5 are used.

\section*{Acknowledgments}
This material is based upon work supported by the National Science Foundation under grants DMR-1905909 and DMR-2330562 at the University of Washington (S.E.), as well as partial support through the University of Washington Materials Research Science and Engineering Center under grant DMR-2308979 (J.L.). We thank Sean Suh for performing some preliminary magnetization measurements.
Work at Seikei University (M.M.) was supported by the Japan Science and Technology Agency (JST) Fusion Oriented Research for disruptive Science and Technology (FOREST; grant No. JPMJFR202G, Japan). A part of this work at Seikei University was supported by Japan Society for the Promotion of Science (JSPS) KAKENHI with No. 23K26147 and No.23H01453. The work at the National Institute of Advanced Industrial Science and Technology (AIST) was supported by NEDO. No competing interests are declared by all authors.

\vspace{0.3 cm}

\section*{Author Contributions}
S.E. and M.M. conceived and designed the experiment.
M.M., A.I. and T.I. grew the (Y,Gd)BCO and EuBCO films.
D.Y. and T. K. performed microstructural studies.
J.L. performed magnetization studies and data analysis.
S.E. determined data analysis procedures, S.E. and J.L. thoroughly reviewed the data analysis.
S.E. and J.L. wrote the manuscript.
All authors commented on the manuscript.

\pagebreak
\let\clearpage\relax 
\onecolumngrid
\section*{Supplemental Materials}\label{sec:smaterials}
\setcounter{figure}{0}

\makeatletter 
\renewcommand{\thefigure}{S\@arabic\c@figure}
\makeatother

\setcounter{figure}{0}
\makeatletter 
\renewcommand{\thefigure}{S\@arabic\c@figure}
\makeatother

\setcounter{table}{0}
\makeatletter 
\renewcommand{\thetable}{S\@arabic\c@table}
\makeatother

\newcommand{\vect}[1]{\mathbf{#1}}
\DeclareSIUnit\oersted{Oe}

\maketitle

\section*{Determining the critical temperature $T_c$} 

To extract the transition temperature $T_c$, we measured the magnetic moment $m$ at a fixed magnetic field $\mu_0H=0.2-0.4$ mT while sweeping the temperature in our Quantum Design MPMS3 magnetometer, as shown in Fig.~\ref{fig:figS1}. Here, it is clearly shown that as the concentration of BaHfO$_3$ (BHO) increases, the midpoint of the transition shifts to a lower temperature. To systematically determine the $T_c$ onset, the moment vs temperature data is first normalized by defining the minimum moment as -1. Next, the average normalized moment for the background signal is calculated and subtracted from the data, such that the average moment becomes 0 at the normal state. The onset $T_c$ is then defined as when the normalized moments are within a threshold of -0.04. The $T_c$ is extracted to be 92.81 K, 91.7 K, 90.5 K, 90.34 K, and 90.18 K for the films containing 0 - 4.8 vol\% BHO, respectively, thereby decreasing with increasing BHO content.

\begin{figure*}[h!]
\centering
\includegraphics[width=0.6\linewidth]{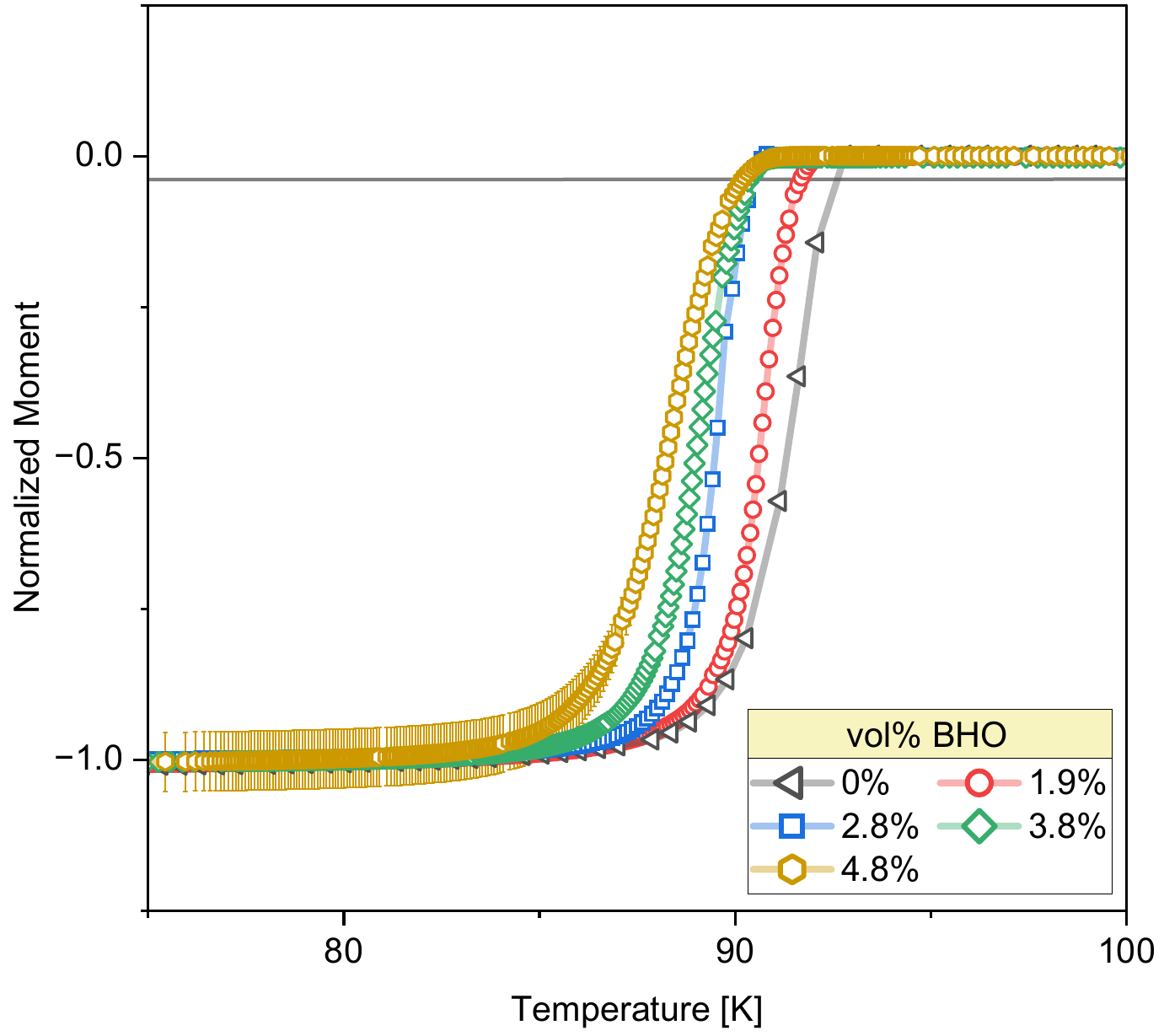}
\caption{\label{fig:figS1} \textbf{Normalized moment versus temperature for EuBCO with different concentrations of BHO.} The applied magnetic field ranges from 0.2 to 0.4 mT. The horizontal line is the -0.04 threshold used to extract $T_c$. The error bar here represents the DC mode moment fix center error measured by MPMS3 systems.}
\end{figure*}

\section*{Chemical composition of EuBCO films}

Figure \ref{fig:figS2} displays EDS images containing elemental maps of the distribution of Eu, Ba, Cu, Hf, and O in the film containing 4.8 vol\% BHO. The cross-sectional (a-f) images reveal that the formed BHO nanorods are aligned, but segmented, and the disconnected region is Eu- and Cu-rich, and Hf-poor. Note that the nanorod density appears higher than in actuality because the cross-section images capture multiple layers, such that nanorods at different depths appear adjacent. The plan-view images (g-l) show that the BHO nanorods in 4.8 vol\% BHO films are approximately 20 nm apart.

\begin{figure*}[h!]
\centering
\includegraphics[width=1\linewidth]{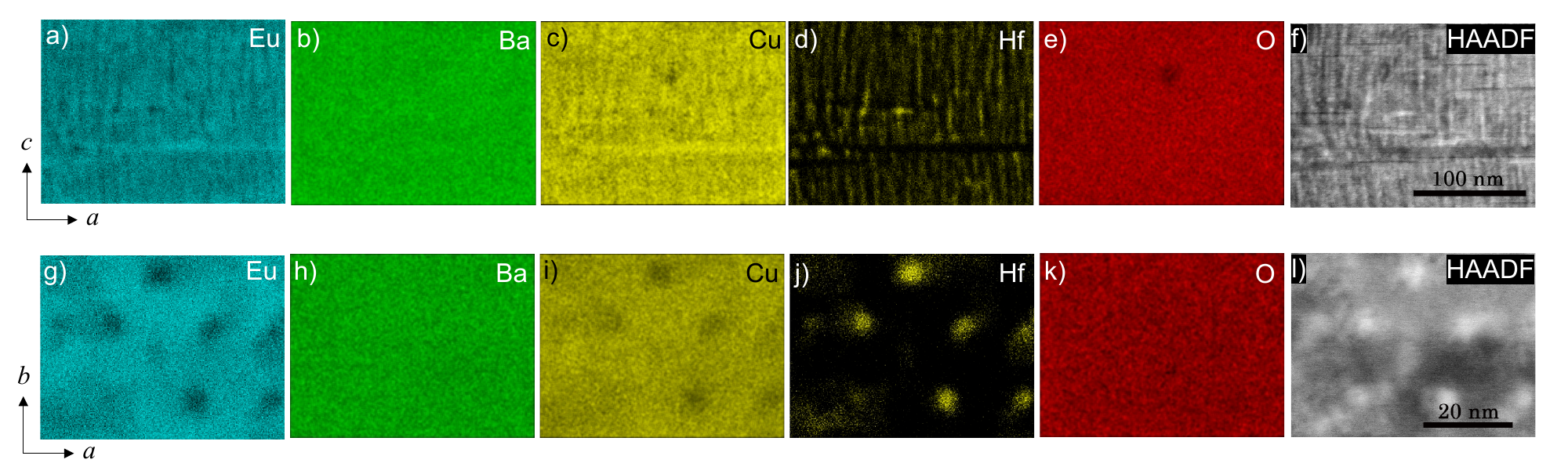}
\caption{\label{fig:figS2} \textbf{Elemental distribution and microstructural characterization of EuBCO film with 4.8 vol\% BHO.} (\textbf{a}-\textbf{f}) Cross-section and (\textbf{g}-\textbf{l}) planar-view of energy dispersive spectroscopy results revealing the chemical composition of the EuBCO film containing 4.8 vol\% BHO. The included elements are (\textbf{a}, \textbf{g}) Europium, (\textbf{b}, \textbf{h}) barium, (\textbf{c}, \textbf{i}) copper, (\textbf{d}, \textbf{j}) hafnium, and (\textbf{e}, \textbf{k}) oxygen. (\textbf{f}, \textbf{l}) High-angle annular dark-field (HAADF) STEM image. }
\end{figure*}

\section*{Magnetic Hysteresis Loops}

\begin{figure*}[h!]
\centering
\includegraphics[width=1\linewidth]{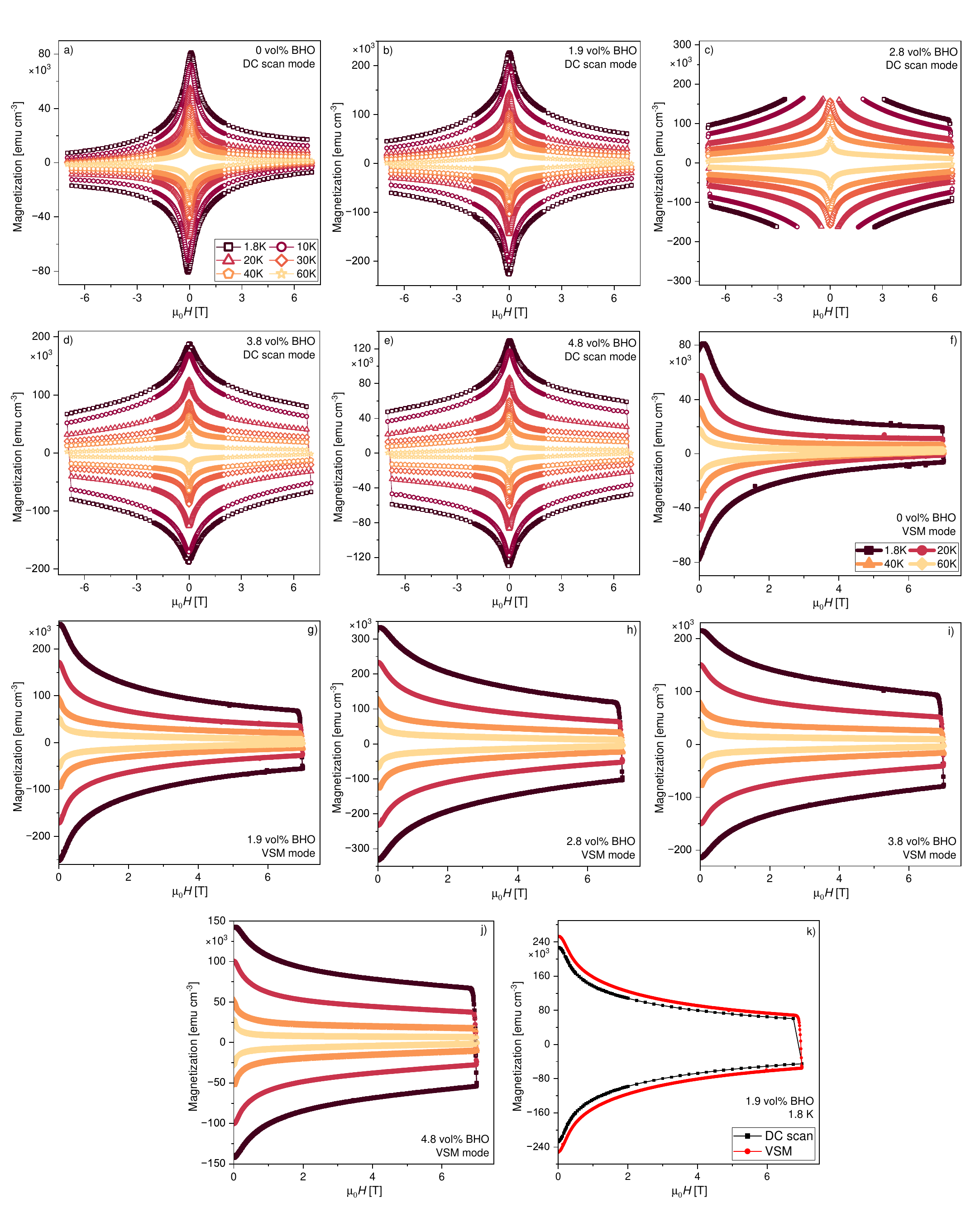}
\caption{\label{fig:figS3} \textbf{Magnetic hysteresis loops collected using DC scan mode and VSM mode for EuBCO films at different temperatures.} (\textbf{a, f}) 0 vol\%, (\textbf{b, g}) 1.9 vol\%, (\textbf{c, h}) 2.8 vol\%, (\textbf{d, i}) 3.8 vol\%, and (\textbf{e, j}) 4.8 vol\% BHO for DC scan mode and VSM mode, respectively. The missing data for 2.8 vol\% in (\textbf{c}) is due to a large (saturated) magnetic moment beyond the sensitivity of our magnetometer. (\textbf{k}) Hysteresis loops collected using VSM mode and DC scan mode for 1.9 vol\% BHO at 1.8 K for comparison.}
\end{figure*}

\begin{figure*}[h!]
\centering
\includegraphics[width=1\linewidth]{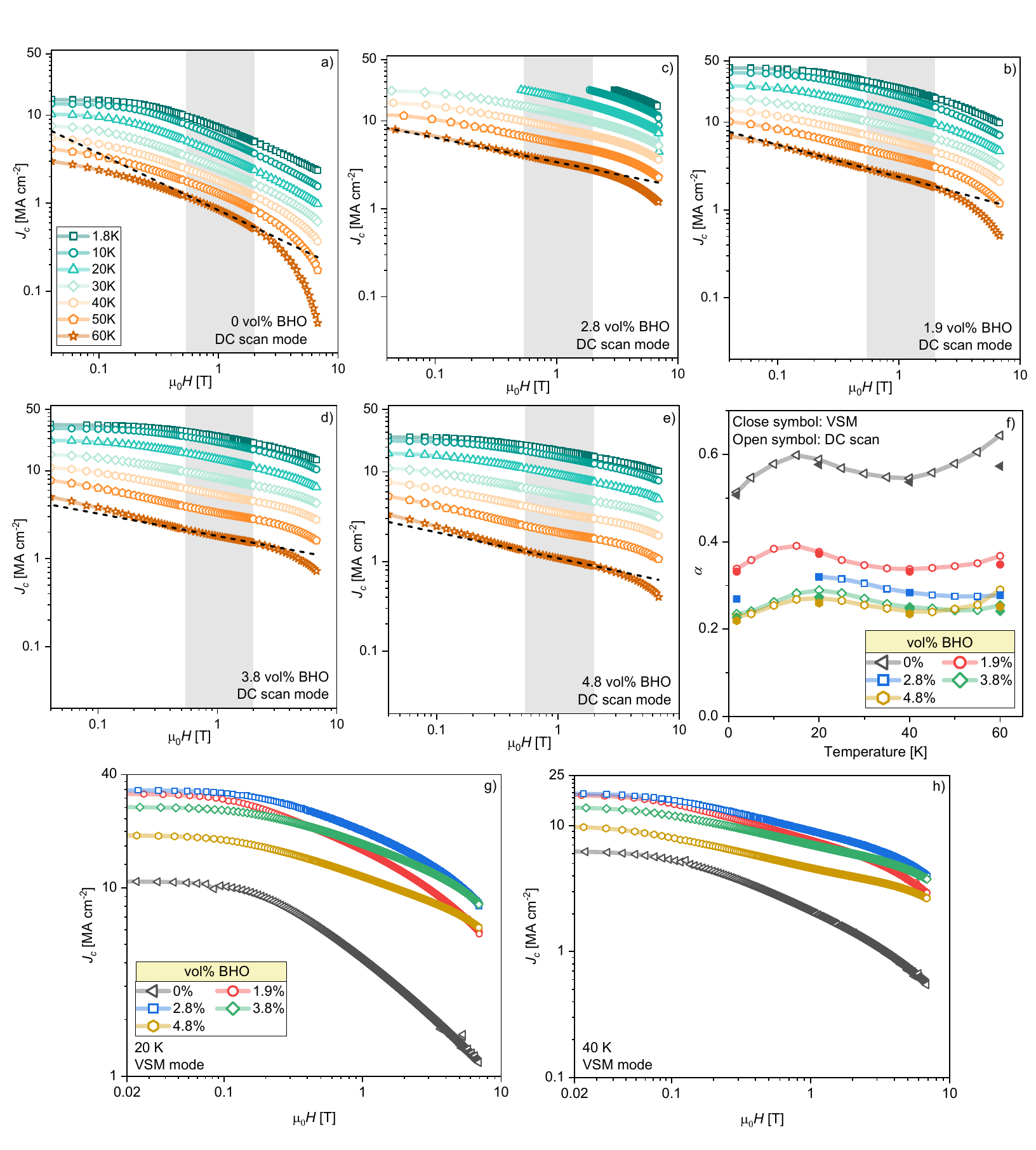}
\caption{\label{fig:figS4} \textbf{DC scan mode and VSM mode field-dependent critical current density $J_c$ at different temperatures for EuBCO films.} DC scan mode: (\textbf{a}) 0 vol\%, (\textbf{b}) 1.9 vol\%, (\textbf{c}) 2.8 vol\%, (\textbf{d}) 3.8 vol\%, and (\textbf{e}) 4.8 vol\% BHO. Here, $J_c$ is calculated using the Bean critical state model from the magnetization data $m(H)$ collected under DC scan mode. The missing data for the sample containing 2.8 vol\% BHO in (\textbf{c}) is due to a large moment beyond what is measurable using DC scanning mode. (\textbf{f}) Exponent $\alpha$ extracted from the field-dependent $J_c \propto B^{-\alpha}$ at different temperatures. The dashed lines are example fits applied within the shaded region (0.55 - 2 T). Exponents extracted from data collected using VSM and DC scan modes are compared. Field-dependent $J_c$ from VSM mode $m(H)$ results for EuBCO films with different BHO doping levels at a temperature of (\textbf{g}) 20 K and (\textbf{h}) 40 K. }
\end{figure*}

Magnetic hysteresis loops were collected for all samples between 1.8 and 75 K using the DC scan mode. Temperature-dependent loops are shown in Fig.~\ref{fig:figS3}(a–e). Note that the missing data for the sample containing 2.8 vol\% BHO is due to the sample's large moment, which exceeds the DC scan mode limit. To resolve this, we employed vibrating sample magnetometry (VSM), which allows for higher moment measurements. For consistency, VSM was also used to measure all other samples at temperatures (1.8, 20, 40, and 60 K) as shown in Figure~\ref{fig:figS3}(f-j). Figure~\ref{fig:figS3}(k) compares the hysteresis loops obtained via DC scan and VSM modes for the 1.9 vol\% BHO sample at 1.8 K. In this case, the DC scan was conducted using a 30 mm scan length and a 4 s scan time at stabilized fields. VSM measurements yielded moments approximately 20\% higher on average. For the 2.8 vol\% BHO sample, measured using a 10 mm scan length and 1 s scan time under continuous sweep mode, the VSM moment was on average 8.5\% higher. This enhancement is attributed to magnetic relaxation, as VSM mode can measure 4–10 times faster than those in DC scan mode, which captures the moment later in the creep process.

Subsequently, the critical current density $J_c$ was calculated using the Bean critical state model from the $m(H)$ data collected in DC scan mode, with representative curves shown in Fig.\ref{fig:figS4}(a–e). Because the main text uses VSM-derived data for the $J_c$ analysis, here we assess whether the scan mode affects the result. Accordingly, Fig.~\ref{fig:figS4}(f) compares the power-law exponent $\alpha(T)$ extracted from both DC scan (open symbols) and VSM (closed symbols) modes. The $\alpha$ values derived from the DC scan data show weak temperature dependence and a clear decreasing trend with increasing BHO content. A similar trend is observed for the VSM-derived $\alpha$. Notably, between 1.8 and 40 K, the VSM and DC scan values of $\alpha$ agree within 5.5\%. For $\alpha$ at 60 K, the VSM data results in lower $\alpha$ than the DC mode data. This discrepancy can be attributed to increased creep at higher temperatures, such that VSM mode may more accurately capture $J_c \propto m$, by measuring $m$ earlier in the magnetic relaxation process. Lastly, $J_c(H)$ obtained from VSM $m(H)$ measurements is shown in Fig.~\ref{fig:figS4}(g,h) for 20 and 40~K, while the 1.8 and 60~K data are included in the main text.

\section*{Magnetic Relaxation}

\begin{figure*}[h!]
\centering
\includegraphics[width=1\linewidth]{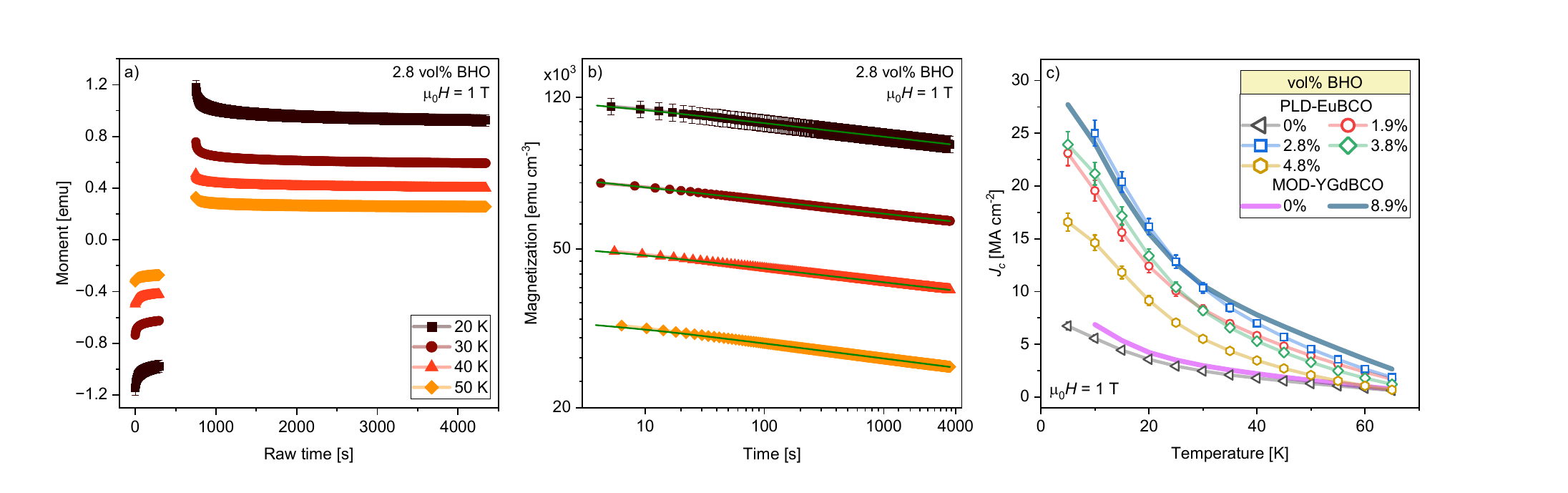}
\caption{\label{fig:figS5} \textbf{Magnetic relaxation processing and derived critical current density using DC scan mode.} (\textbf{a}) Example of the raw magnetic relaxation data $m(t)$ collected on the sample containing 2.8 vol\% BHO at $\mu_0H=1$ T for different temperatures. (\textbf{b}) The same data after background subtraction and adjustment for system delay time $t_d$. Here, $\ln M$ versus $\ln (t+t_d)$ is plotted, and the slope of the linear fit is defined as $S$. (\textbf{c}) Critical current density $J_c$ against temperatures for EuBCO and (Y,Gd)BCO with different vol\% BHO at $\mu_0H=1$ T. Here, $J_c$ is calculated from Bean's critical state model using the first data point from magnetic relaxation measurements. All the error bars here originate from the DC mode moment free center error measured by MPMS3 systems.}
\end{figure*}

\begin{figure*}[h!]
\centering
\includegraphics[width=1\linewidth]{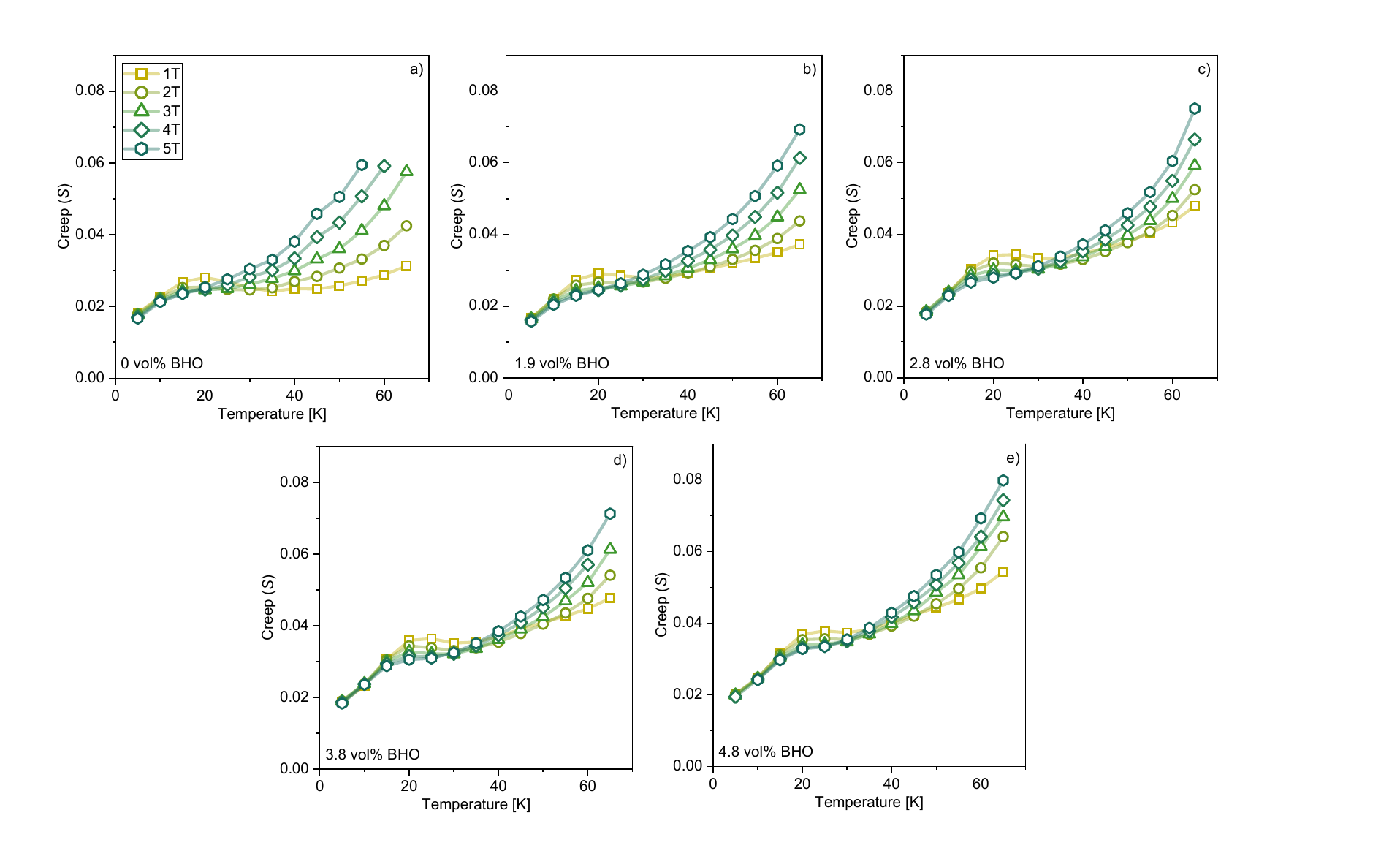}
\caption{\label{fig:figS6} \textbf{Vortex creep parameter $S = -d\ln M/d\ln t$ versus temperatures at different applied magnetic fields for the EuBCO films.} (\textbf{a}) 0 vol\% (\textbf{b}) 1.9 vol\%, (\textbf{c}) 2.8 vol\%, (\textbf{d}) 3.8 vol\%, and (\textbf{e}) 4.8 vol\% BHO. }
\end{figure*}

\begin{figure*}[h!]
\centering
\includegraphics[width=1\linewidth]{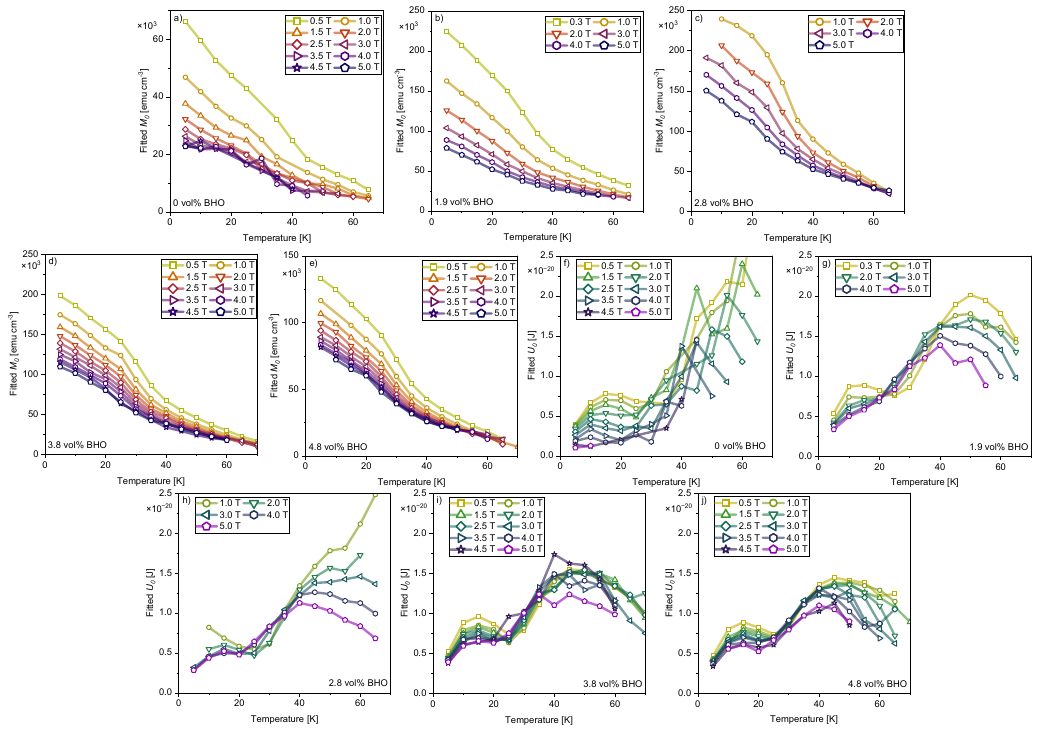}
\caption{\label{fig:figS7} \textbf{Temperature dependence of fitted $M_0$ and $U_0$ at different applied magnetic fields for the EuBCO films.} (\textbf{a, f}) 0 vol\%, (\textbf{b, g}) 1.9 vol\%, (\textbf{c, h}) 2.8 vol\%, (\textbf{d, i}) 3.8 vol\%, (\textbf{e, j}) 4.8 vol\%.}
\end{figure*}

Magnetic relaxation data $m(t)$ were collected using DC scan mode at magnetic fields of 1 - 5 T, with measurements performed at each field over a temperature range of 5 to 75 K. Figure~\ref{fig:figS5}(a) shows the raw relaxation data for the sample with 2.8 vol\% BHO under an applied field of $\mu_0H=1$ T. Magnetic relaxation was measured for 5 minutes on the lower branch and 60 minutes on the upper branch of the hysteresis loop, in both cases after the critical state was established. The background signal from the sample mount was determined by summing the moments from the upper and lower branches, averaging the result, and dividing by two. This background was then subtracted from the upper branch relaxation data. To correct for the time lag between the application of the magnetic field and the initial measurement taken by the MPMS3—during which magnetic relaxation has already begun—and to incorporate the effects of initial flux redistribution, a system delay time $t_d$ is introduced when extracting the vortex creep parameter $S$. Here, $S = -d\ln m / d\ln(t + t_d)$, where $t_d$ is a fitting parameter chosen to maximize the correlation coefficient. On average, $t_d$ is approximately 2 seconds. Lastly, the moment $m(t)$ is converted to magnetization $M(t)$ using the sample dimensions, which are included in Table \ref{tab:samples1}. The final processed magnetization data used to determine $S$ is shown in Fig.~\ref{fig:figS5}(b). Figure~\ref{fig:figS5}(c) presents the temperature dependence of $J_c$ at $\mu_0H=1$ T for all seven samples, calculated using the Bean model based on the first measurement of $m$ in the magnetic relaxation data. As expected, $J_c$ decreases monotonically with increasing temperature. Comparing samples, having 2.8 vol\% BHO produces the highest $J_c$-values at all temperatures for EuBCO films. Figure~\ref{fig:figS6}(a-e) shows the temperature dependence of $S$ at fixed applied magnetic fields for all samples. Below $\sim 25 \mathrm{\ K}$, $S$ is smaller at higher magnetic fields, whereas above it, lower fields result in smaller values of $S$. 
\begin{table}[h!]
\caption{\textbf{Sample dimensions.}}
\label{tab:samples1}
\begin{center}
\begin{tabular*}{0.9\textwidth}{@{\extracolsep{\fill}}ccccccc}
\hline\hline
sample & vol. \% & length $l$ & width $w$ & thickness $\delta$ \\
ID & BHO & mm & mm & nm \\
\hline
1 & 0   & 3.20 & 3.04 & 600 & \\
2 & 1.9 & 3.31 & 3.21 & 550 & \\
3 & 2.8 & 4.46 & 4.05 & 550 & \\
4 & 3.8 & 3.43 & 3.30 & 550 & \\
5 & 4.8 & 3.20 & 3.14 & 550 & \\
\hline\hline
\end{tabular*}
\end{center}
\end{table}

Next, to extract the glass exponent $\mu$, we fitted our background subtracted magnetic relaxation data $M(t)$ using
\[
M(t) = M_{0} \left\{ 1 + \frac{\mu k_B T \ln[(t + t_d)/t_0]}{U_0} \right\}^{-1/\mu},
\]
as given in Eq.~(3) of the main text. To minimize the number of free parameters, we fixed $t_d = 2$~s and $t_0 = 10^{-6}$~s, while allowing $U_0$, $M_0$, and $\mu$ to vary. Figure~\ref{fig:figS7}(a–j) presents the fitted $U_0$ and $M_0$ values at different temperatures and applied fields. Here, the fitted $M_0$ decreases monotonically with increasing temperature, as expected, while the fitted $U_0$ remains mostly temperature independent between 10 and 35 K and becomes temperature dependent at higher temperatures. 

Last, we investigate how the choice of $t_0$ influences the extracted fit parameters. We varied $t_0$ from $10^{-7}$ to $1$~s to extract the temperature-dependent $M_0$, $U_0$, and $\mu$, as shown in Fig.~\ref{fig:figS9}. The temperature dependencies of $M_0$ and $U_0$ are sensitive to the choice of $t_0$. However, the extracted glassy exponent $\mu$ is unaffected by the value of $t_0$. The curves completely overlap in Fig.~\ref{fig:figS9}(c) because the extracted $\mu$ values differ by less than $10^{-4}$. Thus, although $t_0$ influences $M_0$ and $U_0$, the determination of $\mu$ is robust against the choice of $t_0$.

\section*{Comparison of EuBCO and YBCO Nanocomposites with B\textit{M}O Nanorods}

Figure S8(a) shows that the field-dependent $J_c$ at 65 K increases with the addition of B\textit{M}O for both EuBCO and YBCO. At the same time, the vortex creep rate $S$ also increases, as shown in Figure S8(b). Here, the EuBCO data is from the current study, where 5 and 7.5 mol\% produce 2.8 and 3.8 vol\% BHO, respectively. The YBCO data are taken from Maiorov \textit{et al.} \textit{Nat. Mater.} \textbf{8}, 398 (2009). and Mele \textit{et al.} \textit{Supercond. Sci. Technol.} \textbf{21}, 032002 (2008), where the closest available growth conditions, applied fields, and measurement temperatures to our samples were used for comparison. The film reported by Maiorov \textit{et al.} was grown by PLD with 5 mol\% BZO at an oxygen partial pressure of 200 mTorr and a deposition temperature of $830\,^{\circ}$C. The film reported by Mele \textit{et al.} was grown using PLD with 4 wt\% BZO under an oxygen partial pressure of 200 mTorr and a substrate temperature of $800\,^{\circ}$C.

\begin{figure*}[h!]
\centering
\includegraphics[width=0.9\linewidth]{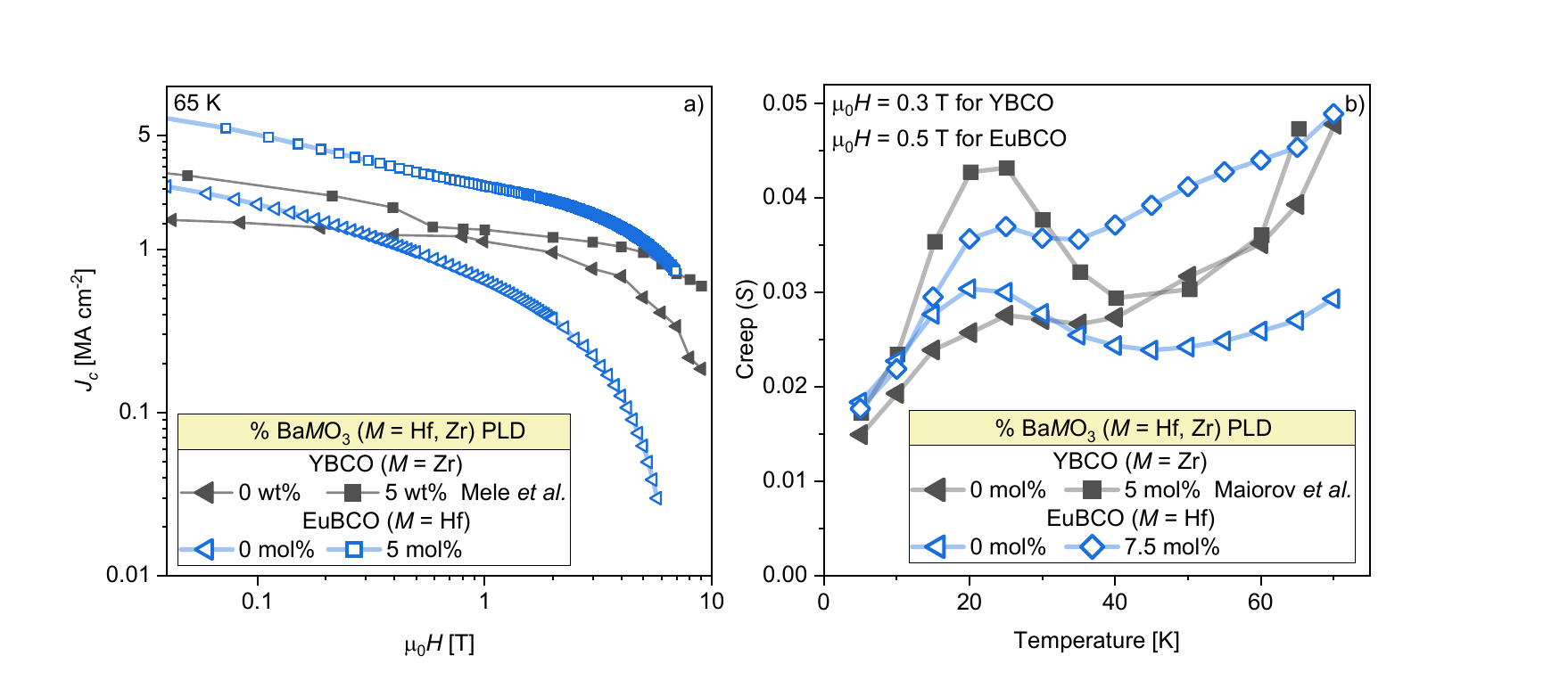}
\caption{\label{fig:figS8} \textbf{Comparison of PLD-grown YBCO and EuBCO films with and without B\textit{M}O nanorods.} The YBCO data are taken from \textit{Nat. Mater.} \textbf{8}, 398 (2009). and \textit{Supercond. Sci. Technol.} \textbf{21}, 032002 (2008), while the EuBCO data are from this work. (a) Field-dependent critical current density $J_c$ at 65 K. (b) Vortex creep rate $S = -d\ln M/d\ln t$ as a function of temperature. For consistency, the closest available applied fields were chosen: $\mu_0 H = 0.3$ T for YBCO and $\mu_0H = 0.5$ T for EuBCO.}
\end{figure*}

\begin{figure*}[h!]
\centering
\includegraphics[width=1\linewidth]{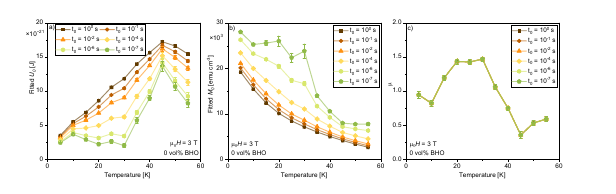}
\caption{\label{fig:figS9} \textbf{Temperature dependence of fitted $U_0$, $M_0$, and $\mu$ using different $t_0$.} (\textbf{a}) Temperature-dependent $U_0$, (\textbf{b}) temperature-dependent $M_0$, and (\textbf{c})  temperature-dependent $\mu$. The examples are chosen from EuBCO without BHO at $\mu_0 H = 3$ T. In (c), the curves completely overlap because the extracted $\mu$ values differ by less than $10^{-4}$.}
\end{figure*}

\end{document}